\RequirePackage{letltxmacro}
\LetLtxMacro{\LaTeXtextbf}{\textbf}
\documentclass{ieeeaccess}
\LetLtxMacro{\textbf}{\LaTeXtextbf}
\usepackage{cite}
\usepackage{amsmath,amssymb,amsfonts}
\usepackage{graphicx}
\usepackage[caption=false]{subfig}

\usepackage{textcomp}
\usepackage{algorithmic}
\usepackage[linesnumbered,ruled,vlined]{algorithm2e} 
\usepackage{mathtools}

\SetCommentSty{mycommfont}
\SetKwInput{KwInput}{Input}
\SetKwInput{KwOutput}{Output}
\usepackage{bm}

\usepackage{bm}
\makeatletter
\AtBeginDocument{\DeclareMathVersion{bold}
\SetSymbolFont{operators}{bold}{T1}{times}{b}{n}
\SetSymbolFont{NewLetters}{bold}{T1}{times}{b}{it}
\SetMathAlphabet{\mathrm}{bold}{T1}{times}{b}{n}
\SetMathAlphabet{\mathit}{bold}{T1}{times}{b}{it}
\SetMathAlphabet{\mathbf}{bold}{T1}{times}{b}{n}
\SetMathAlphabet{\mathtt}{bold}{OT1}{pcr}{b}{n}
\SetSymbolFont{symbols}{bold}{OMS}{cmsy}{b}{n}
\renewcommand\boldmath{\@nomath\boldmath\mathversion{bold}}}
\makeatother

\def\BibTeX{{\rm B\kern-.05em{\sc i\kern-.025em b}\kern-.08em
    T\kern-.1667em\lower.7ex\hbox{E}\kern-.125emX}}

\begin{document}
\history{Date of publication xxxx 00, 0000, date of current version xxxx 00, 0000.}
\doi{10.1109/ACCESS.2023.1120000}

\title{Traffic and Obstacle-aware UAV Positioning in Urban Environments Using Reinforcement Learning}
\author{\uppercase{Kamran Shafafi}\authorrefmark{1}, \uppercase{Manuel Ricardo}\authorrefmark{1}, \IEEEmembership{Member, IEEE}, and \uppercase{ Rui Campos}\authorrefmark{1}, \IEEEmembership{Senior Member, IEEE}}
\address[1]{INESC TEC and Faculty of Engineering, University of Porto, Porto, Portugal}


\tfootnote{This work is financed by National Funds through the Portuguese funding agency, FCT – Fundação para a Ciência e a Tecnologia, under the PhD grant 2023.00384.BD.}

\markboth
{Author \headeretal: Preparation of Papers for IEEE TRANSACTIONS and JOURNALS}
{Author \headeretal: Preparation of Papers for IEEE TRANSACTIONS and JOURNALS}

\corresp{Corresponding author: First A. Author (e-mail: kamran.shafafi@inesctec.pt).}

\begin{abstract}
Unmanned Aerial Vehicles (UAVs) are suited as cost-effective and adaptable platforms for carrying Wi-Fi Access Points (APs) and cellular Base Stations (BSs). Implementing aerial networks in disaster management scenarios and crowded areas can effectively enhance Quality of Service (QoS). In such environments, maintaining Line-of-Sight (LoS), especially at higher frequencies, is crucial for ensuring reliable communication networks with high capacity, particularly in environments with obstacles. The main contribution of this paper is a traffic- and obstacle-aware UAV positioning algorithm named Reinforcement Learning-based Traffic and Obstacle-aware Positioning Algorithm (RLTOPA), for such environments. RLTOPA determines the optimal position of the UAV by considering the positions of ground users, the coordinates of obstacles, and the traffic demands of users. This positioning aims to maximize QoS in terms of throughput by ensuring optimal LoS between ground users and the UAV. The network performance of the proposed solution, characterized in terms of mean delay and throughput, was evaluated using the ns-3 simulator. The results show up to 95\% improvement in aggregate throughput and 71\% in delay without compromising fairness.
\end{abstract}

\begin{keywords}
Unmanned Aerial Vehicles, UAV positioning, aerial networks, LoS communications technology, reinforcement learning, high-capacity communications, positioning algorithms.
\end{keywords}

\titlepgskip=-15pt
\maketitle

\section{Introduction}
\label{sec:introduction}
\PARstart{U}{nmanned} Aerial Vehicles (UAVs) are used in various applications, namely agriculture, inspections, security missions, and industries. These capabilities have led to significant interest from the research community in recent years \cite{GUPTA2022154291}. UAVs are highly mobile, making them ideal for rapid deployment in emergency scenarios, including cyberattacks and terrorist threats, as well as natural disasters like wildfires, earthquakes, and floods \cite{shafafi2023joint}. Their dynamic capabilities and ability to swiftly adapt to changing network conditions and environments enhance their utility \cite{Xiao2021ASO}. In telecommunications networks, aerial platforms, like balloons, high-altitude platforms (HAPs), and UAVs, play a crucial role in boosting the capacity of terrestrial cellular networks. These platforms are especially valuable for meeting the evolving requirements of next-generation wireless networks due to their cost-effectiveness, enhanced mobility, hovering capability, and adjustable altitude \cite{10028754}. UAVs are particularly well-suited for carrying Base Stations (BSs) and Wi-Fi Access Points (APs) due to their adaptability \cite{shafafi2023uav}.

\Figure[t!](topskip=0pt, botskip=0pt, midskip=0pt){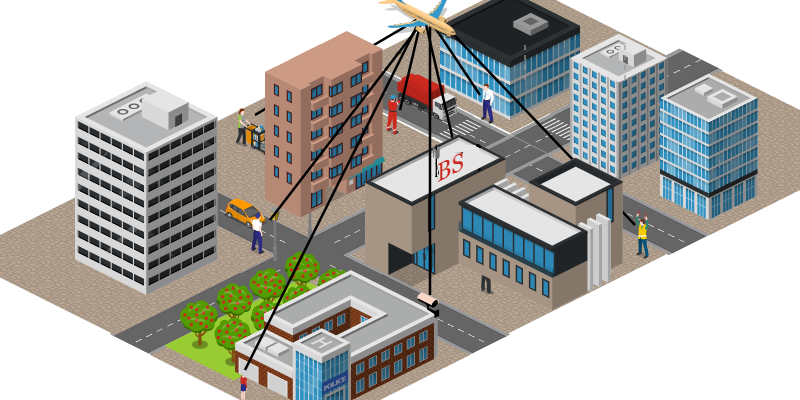}
{ \textbf{Urban environment consisting of UAV located in a 3D position as FAP to provide LoS wireless connectivity to users. The UAV is then connected to the LTE BS.}\label{fig1}}

A significant challenge in deploying Flying Access Points (FAPs) is optimizing their positioning to meet user demands, thereby enhancing network capacity and coverage, increasing surveillance security, and managing disaster scenarios \cite{articleElnabty, articlePogaku}. In modern wireless systems—especially in public safety, Intelligent Transportation Systems (ITS), Internet of Things (IoT) systems, and smart cities—precise positioning data is of paramount importance \cite{f8048d2d1f9d4a9bb4638d94d148e74e, articlechen}. Figure \ref{fig1} illustrates an urban environment where a fixed BS is located on the rooftop but cannot fully serve all User Equipment (UE) due to obstacles obstructing the Line-of-Sight (LoS), especially at higher frequencies. 

By ensuring LoS between the UAV and the UE, the maximum capacity of the links can be achieved, accommodating the traffic demands even in obstacle-rich environments. The optimal position for the UAV is the position that leads to the highest number of UEs with wireless connectivity and maximum aggregate throughput. UAV placement has been extensively studied \cite{8377408, 9448966, 8422376, 8761897, 7962642, ML8121867, MLJiang2017MachineLP }, but existing solutions mainly focus on obstacle-free scenarios or propagation loss models for obstacle-rich scenarios without addressing the UAV positioning problem. 

The main contribution of this paper is a traffic and obstacle-aware UAV positioning algorithm for obstacle-rich environments, named RLTOPA. RLTOPA takes into account the positions of ground users, obstacles, and users’ traffic demands to determine the optimal UAV position. This approach aims to maximize Quality of Service (QoS), in terms of throughput, by maintaining LoS between the UAV and ground users. We use Reinforcement Learning (RL) to solve the problem and determine the optimal position of the UAV. The proposed RLTOPA algorithm was evaluated with the Network Simulator 3 (ns-3) to assess network performance based on the UAV positioning defined by RLTOPA.

The rest of this paper is organized as follows. Section II provides an overview of the related works. Section III introduces the system model. Section IV formulates the problem. Section V presents RLTOPA, including its underlying rationale and its use in a simple scenario. Section VI focuses on the performance evaluation, including the simulation setup scenarios, performance metrics, and simulation results. Finally, Section VII outlines the main conclusions and the potential future work directions.

\section{Related Works}
\label{sec:relatedwork}

There is a wide range of published research addressing the UAV positioning problem. In this section, we discuss the most relevant works and their limitations, providing insights into how our work aims to address and improve upon them.

Many studies have considered obstacle-free environments for UAV positioning, primarily focusing on deploying UAVs as relays between the Radio Access Network (RAN) and a backhaul network. The authors in \cite{8377408} proposed traffic-aware multi-tier flying network planning to improve throughput in temporary events without obstacles. A positioning algorithm for an obstacle-free backhaul network is proposed in \cite{9448966}. The authors in \cite{8422376, 8761897, 7962642} have suggested ways to increase the aerial network's service area. Positioning solutions in \cite{8038014, 7738405, 7572068} aim to improve the QoS and Quality of Experience (QoE) for UE. In \cite{LIU2023103047} is proposed a deployment strategy to ensure the QoS and minimize interference in obstacle-free environments using minimal UAVs. In recent years, researchers have developed positioning algorithms using Machine Learning (ML) techniques \cite{ML8121867, MLJiang2017MachineLP}. \cite{43-bab6dcd317da4e3d82375669cbb45023}, utilizes deep learning to maximize the user throughput. The authors in \cite{23-8644345} proposed a three-step method for 3D deployment and dynamic movement of multiple UAVs. Enhancing QoS and QoE through Q-learning techniques is the main contribution of \cite{21-8377340}, and \cite{22-COLONNESE2019101872}.  

Several studies have addressed the problem of potential obstacles that UE may face when encountering obstructions. In \cite{9681949}, the authors propose a comprehensive system that utilizes point cloud processing to predict blockages for UE. \cite{9512383} employs video-based techniques to proactively anticipate dynamic link failures between a 5G BS and the user. Furthermore, \cite{9464922} leverages computer vision techniques to predict blockage scenarios in millimeter-wave channels, enabling proactive measures before the blockage occurs. However, these studies do not consider the positioning of UAVs in FNs and instead focus solely on providing wireless directions through terrestrial BSs.

Various studies have made valuable contributions to environments with obstacles, but these works did not focus on positioning approaches. They have proposed propagation loss models for obstacle-rich environments or aimed to reduce transmission power \cite{7921981, 8316776}. Specifically, \cite{8352733} introduced a propagation loss model for indoor and outdoor settings in the lower Super High-Frequency (low-SHF) band. \cite{wu2021delay} presented an algorithm to minimize delay in UAV positioning. In \cite{8450437}, the focus was on determining the optimal placement of a single UAV to maximize the duration of uplink transmissions. \cite{Hayajneh20213DDO} investigated the impact of obstacles on UE and calculated the probability of LoS. It also explored using UAVs to support terrestrial BS in serving the UE. For mmWave NR deployments in urban areas, \cite{s22030977} proposed a distinct propagation loss model to estimate LoS blockage probability. Resource allocation in multi-UAV scenarios was addressed through an optimization problem formulated by \cite{10199180, 10210623}, aiming to maximize the achievable rate of UE by allocating power and associating it with the appropriate UAVs. Several previous studies have also focused on enhancing the accuracy of the loss model in urban settings \cite{111114483593, 22227037248, 33331683399, 444446863654}.

In recent studies, UAV deployment solutions specifically designed for disaster scenarios have been presented \cite{co19453853, co29175054}. These studies primarily focus on achieving maximum coverage and ensuring connectivity for UE. However, they do not address user demand accommodation and obstacles. \cite{8736350} addressed the coverage problem using a combination of RL and a genetic algorithm-based K-means (GAK-means) approach. The authors of \cite{8727504} attempted to optimize UAV trajectories with a Q-learning approach for maximizing the total transmission rate while satisfying the UE rate requirements. Finally, \cite{9312102} proposed a solution to reduce interference in both uplink and downlink when multiple UAVs are deployed.

Previous studies have predominantly focused on either obstacle-free scenarios or obstacle-rich environments, involving terrestrial networks. However, approaches that ignore obstacles are not effective in real-world settings. On the other hand, existing solutions for obstacle-rich environments often lack the provision of high-capacity links through LoS for communication at high-frequencies. For example, the algorithm proposed in \cite{shafafi2023joint} is limited to scenarios with a single obstacle and cannot operate effectively in environments with multiple obstructions. To address these limitations comprehensively, this paper introduces a novel UAV positioning approach tailored for environments with multiple obstacles.

\section{SYSTEM MODEL}
\label{sec:SYSTEMMODEL}

We consider a square area with dimensions of \(S_{venue} \times S_{venue}\), representing the area in which UE may be located, where \(S_{venue}\) is the length of the venue's side in meters. The system is modeled under the International Telecommunication Union (ITU) path loss models \cite{RRECP14149:online} and can fall into urban, suburban, residential, and rural categories. It consists of the following components: i) a UAV carrying an AP acting as a FAP, which is positioned to provide wireless connectivity to the UE, where \(Z_p\) represents the possible volume where the UAV can be positioned; ii) $N$ UE randomly distributed throughout the venue; iii) $M$ buildings, each with different dimensions and altitudes, acting as obstacles. The distances between the buildings vary, modeling a real cityscape with streets and alleys, as depicted in Figure \ref{fig1}; iv) the Processing System (PS), which can be located in the cloud or at an edge node. The objective is to utilize air-to-ground radio links that are short-range and highly directional. This paper specifically concentrates on the single-UAV scenario; scenarios with multiple UAVs will be explored in future work.

The PS undertakes several key responsibilities. Initially, it gathers the coordinates of all UE on the ground by utilizing a Position Recognition Algorithm (PRA) that employs, for instance, a robust Random Forest classifier such as the one described in \cite{article1234}. Secondly, it retrieves the coordinates of obstacles from a predefined table or map of the venue. In this work, only buildings are considered as obstacles; the consideration of moving obstacles such as cars is left for future work. Lastly, by running the RLTOPA algorithm and utilizing the gathered information, the PS determines the optimal position of the UAV. The PS communicates the resulting optimal position to the UAV, which then adjusts its position accordingly.

\section{problem formulation}
\label{sec:formulation}

An aerial network is represented by a directed graph, $G(t_k)=(U, L(t_k))$. Let us consider \(t_k\) as a static time snapshot of the network, where \(t_k = k \times \Delta t, k \in \mathbb{N}_0\) and \(\Delta t \in \mathbb{R}\). In each instant \(t_k\), the positions of all UE remain constant; in the next instant, they will be updated. $U$ contains $N+1$ nodes forming the network, consisting of all UE and the UAV. \(UE_i\) represents the UE number \(i\), where \(i \in \{1, ..., N\}\), and \(L_i(t_k)\) illustrates the link between \(UE_i\) and the UAV. \(UE_i\) transmits \(B_i(t_k)\) bit/s in each \(t_k\) toward the UAV through \(L_i(t_k)\). 

UAV receives \(R_i(t_k)\) bit/s in each \(t_k\) from the $UE_i$ through \(L_i(t_k)\). Accommodating \(B_i(t_k)\) as demanded traffic bit rate by \(UE_i\), requires the minimum link capacity \(C_i(t_k)\) in bit/s, where the maximum channel capacity is $C^{MAX}$ bit/s. Since the wireless medium is shared, we assume that the UAV can listen to all UE. Because of this, the Medium Access Control (MAC) layer uses Carrier Sense Multiple Access with Collision Avoidance (CSMA/CA) mechanisms to prevent network packet collisions by allowing transmissions only when the channel appears to be idle.

Let us define \(P = (x, y, z)\) as the coordinates of the UAV, and \(P_i = (x_i, y_i, z_i)\) as the coordinates of \(UE_i\). We aim to determine the optimal position of the UAV, \(P\), at any \(t_k\), ensuring the network can accommodate the \(B_i(t_k)\) bit/s by establishing LoS connectivity with all UE. The following rationale is considered in this computation: it is possible to ensure the capacity \(C_{i}(t_k)\) is adequate to accommodate \(B_i(t_k)\) given the transmission power \(P_T\) of each UE, where the total network capacity \(C(t_k) = \sum_{i=1}^{N}C_{i}(t_k)\). When \(C_i(t_k)\) is adequate to accommodate \(B_i(t_k)\), the bitrate received by the UAV is maximized. The coordinates \((x^{MIN}, y^{MIN}, z^{MIN})\) and \((x^{MAX}, y^{MAX}, z^{MAX})\) define the minimum and maximum allowable coordinates based on \(P_T\) and the demanded traffic, where the UAV can be deployed.

Additionally, let $LoS_i$ represent a binary variable that defines whether there is LoS between $UE_i$ and the UAV. Then the problem can be formulated as:

\begin{small}
\begin{subequations}\label{objectives}    
    \begin{alignat}{8}
            & \!\underset{(x, y, z)}{\textrm{maximize}}~&   & ~\text{Throughput}= \sum_{i=1}^{N} R_i(t_k)\hspace{1.1cm} i\in \{1,...,N\}\label{eq:objective-function1}\\     
		  &     &     &~C(t_k) \leq C^{MAX}\label{eq:constraint1}\\
            &     &     &~0 < B_i(t_k) \leq C_{i}(t_k)\hspace{1.8cm} i \in \{1,...,N\}\label{eq:constraint2}\\
		  &     &     &~x^{MIN} ~\leq~ x~\leq~ x^{MAX}\hspace{1.7cm}i \in \{1,...,N\}\label{eq:constraint3} \\
             &     &     &~y^{MIN} ~\leq~ y ~\leq~ y^{MAX}\hspace{1.75cm}i \in \{1,...,N\}\label{eq:constraint4} \\
             &     &     &~z^{MIN} ~\leq~ z ~\leq~ z^{MAX}\hspace{1.8cm}i \in \{1,...,N\}\label{eq:constraint5} \\         
             &    &     & \sum_{i=1}^{N} LoS_i = N \hspace{2.5cm} i \in \{1, ..., N\}\label{eq:constraint6}     
	\end{alignat}    
\end{subequations}
\end{small}

\section{RLTOPA Algorithm}
\label{sec:RLTOPA}

Given the complexity of UAV positioning in obstacle-rich environments, we have formulated the problem within the framework of a standard RL approach. RL excels at balancing exploration and exploitation, learning from interactions, generalizing across tasks, and adapting to real-time decision-making. This makes it well-suited for high-dimensional, uncertain, and evolving scenarios. RLTOPA uses the UAV as an agent to learn the optimal policy for positioning itself within the environment.

In the RLTOPA algorithm, the UAV acts as the agent, responsible for learning the optimal policy for actions within the environment based on given states. These states are defined by a set of observations, including the positions of ground users, obstacles, and traffic demands. After executing an action in a specific state, the environment provides a reward to the agent reflecting the effectiveness of that action. To address the problem, we employ a Deep Q-learning approach to solve the problem using a Deep Q-Network (DQN) algorithm to train the UAV agent. 

In a typical RL scenario, the agent undergoes training and evaluation across discrete episodes, each segmented into a series of time slots ($T \times t_k$). At each time slot \(t_k\), also known as the decision time, the agent collects observations \(obs[t_k]\) from the environment, selects and executes an action \(Act[t_k]\), and receives a reward \(Rew[t_k]\) from the environment. The objective is to learn and evaluate the optimal policy of actions for the agent to maximize the cumulative reward, thereby determining the observations \(obs[t_k]\) that yield the highest \(Rew[t_k]\). The following subsections provide detailed explanations of the RL components, including the observations, actions, and reward function based on the problem formulation. Additionally, the DRL block diagram used in RLTOPA is described.

\subsection{Action space}
\label{sec:action}
In RLTOPA, the action space of the UAV agent is a one-dimensional discrete set of integers, representing the sequential movements of the UAV. The possible positions the agent can take are represented within a 3D Cartesian coordinate grid cube, denoted as $Z_p$. A predefined $gridSize$ determines the step size for UAV movements from one point to the next, enabling systematic scanning of the entire volume $Z_p$. During each episode, the UAV selects a sequence of pre-determined movements, executing one movement per timeslot $t_k$. At each decision interval, the UAV can stay in its current position or move in one of six directions: up, down, left, right, forward, and backward, each corresponding to a movement equal to \emph{gridSize}. This flexibility allows the UAV to effectively approach the target positioning area while retaining the ability to adjust its position as needed.

\subsection{Observation space}
\label{sec:observation}
The main objective of RLTOPA is to find the UAV position that maximizes the QoS in terms of throughput while accommodating of the traffic demand of the UE. To achieve this, the environment must capture a snapshot of its state, gather relevant metrics, and communicate them to the agent. The agent then uses these observations from the previous time interval $t_k$ to make decisions and execute new actions in each decision interval. These observations are crucial for guiding the agent toward decisions that align with the overall objective. The observation space is defined by: i) the 3D coordinates of the UE; ii) the coordinates of the UAV; iii) the number of UE in LoS with the UAV in the timeslot $t_k$.

\subsection{Reward Function}
\label{sec:Ewward}

In RL, the reward function quantifies the desirability of the executed action, guiding the agent to learn a policy \( \pi(Act,s) \) that maximizes the cumulative reward \( Rew[t_k] \) over time. Here \( s \) represents the current state of the environment and \( Act \) is the action taken by the agent at timeslot \( t_k \).

We first consider several key parameters to define \( Rew[t_k] \) in RLTOPA. Let $SNR_i$ in $dB$ represent the Signal-to-Noise Ratio (SNR) of the link between $UE_i$ and the UAV, $P_{T,i}$ in $dBm$ be the transmission power of $UE_i$, and $P_N$ in $dBm$ be the noise floor power. The variable $f$ refers to the frequency of the Wi-Fi technology used, and $d_{max_i}$ is the maximum distance between $UE_i$ and UAV. By enabling the utilization of the Modulation and Coding Scheme (MCS) index \cite{MCSTable25:online}, denoted as $MCS_i$, we select the desired $MCS_i$ based on the user's traffic demand. Then we extract the minimum required $SNR_i$ from this table and use Equation \eqref{eq:friis-propagation-model} to calculate $d_{max,i}$.

\begin{equation}
    \begin{aligned}
        {SNR}_{i} &= P_{T,i} - 20\log_{10}(d_{\text{max,i}}) - 20\log_{10}(f) \\
        &\quad - 20\log_{10}\left(\frac{4\pi}{c}\right) - P_N
    \end{aligned}
    \label{eq:friis-propagation-model}
\end{equation}

Each $UE_i$ is positioned at $(x_i, y_i, z_i)$, defining the center of a sphere with a radius of $d_{max_i}$ representing the transmission range. Within this sphere, the provisioning of $C_{i}(t_k)$ and the fulfillment of the traffic demand $B_i(t_k)$ based on the chosen $MCS_i$ is ensured (Equation \eqref{eq:constraint2}). this means that the UAV's position $P = (x, y, z)$ must lie within this sphere to guarantee the accommodation of the traffic demand. Assuming $d_i$ represents the distance between the UAV and $UE_i$, then:

\begin{equation} 
    \begin{aligned}
        &d_i~\leqslant~d_{max,i}\hspace{2.5cm} i \in \{1,...,N\} \\
        &(x - x_i)^2 + (y - y_i)^2 + (z - z_i)^2 \leqslant \left(10^{\frac{K + P_T - SNR_i}{20}}\right)^2  \\
        &K = -20\log_{10}(f) - 20\log_{10}\left(\frac{4\pi}{c}\right) - P_N
    \end{aligned}
    \label{eq:placement-equations-system}
\end{equation}\vspace{0.1cm}

\begin{figure}
	\centering
		\includegraphics[width=\linewidth]{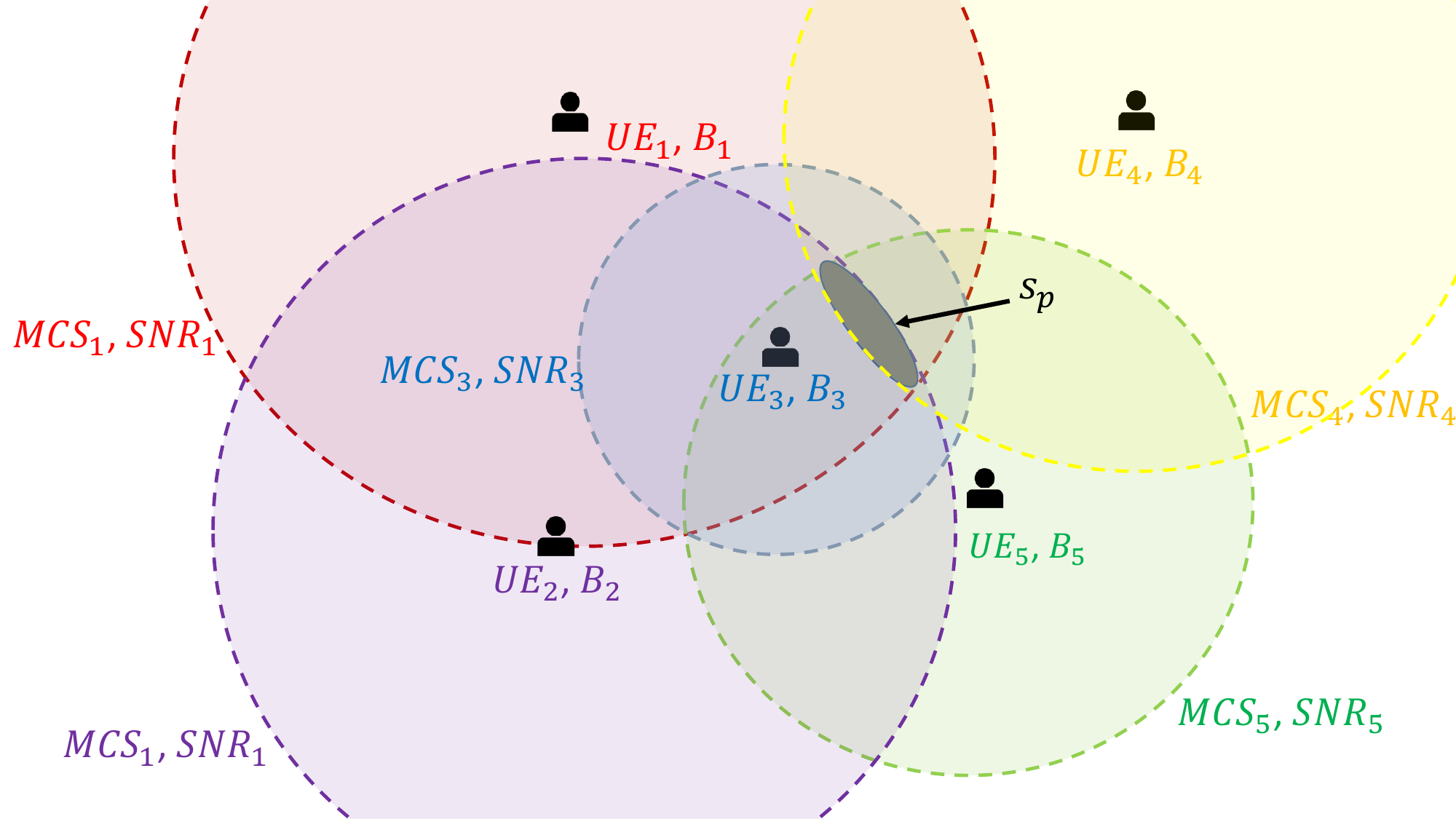}
	\caption{A simple scenario with a two-dimensional (2D) representation of the target positioning subspace ($S_p$) is illustrated. This subspace is determined by the intersection of spheres around each $UE_i$.}
	\label{fig.2}
\end{figure}

Figure \ref{fig.2}, illustrates a straightforward scenario with five UE and different traffic demands $B_i(t_k)$, $i\in \{0,...,N\}$. The gray area in Figure \ref{fig.2} indicates that, to guarantee the accommodation of $B_i(t_k)$, the UAV's position $P$ must be within the joint area of the spheres to satisfy all the traffic demands of the UE (Equations \eqref{eq:constraint3}, \eqref{eq:constraint4}, and \eqref{eq:constraint5}). This space, denoted as $S_p$, emerges as the target location for positioning the UAV. $S_p$ is an optimistic approach since it considers $Friis$ path losses which are smaller than the losses defined by the ITU-T model considered in this work. The ITU-T propagation loss model is a useful model to estimate and predict signal attenuation in urban and suburban. Due to the obstacles, this attenuation is greater than free space. By solving Equation \eqref{eq:placement-equations-system}, derived from Equation \eqref{eq:friis-propagation-model} and considering all UE, a set of potential positions of the UAV ($P$) within the gray area can be calculated.

In obstacle-rich environments, wireless links between the UAV and UE may be obstructed by obstacles, meaning not all positions within the $S_p$ are suitable for UAV positioning. To ensure reliable broadband links that accommodate traffic demands, particularly at higher frequencies, it is necessary to establish LoS between the UAV and all UE. RLTOPA enforces constraints to ensure LoS connectivity resulting in the refined positioning subspace $S_p$. By considering the maximum allowable distance between $UE_i$ and the UAV, $d_{max,i}$, and establishing LoS, maximum network capacity can be achieved to accommodate the traffic demands, thereby maximizing QoS in terms of throughput. Thus, positions that maintain LoS with $UE_i$ are optimal for UAV deployment. 

Within $S_p$, RLTOPA explores positions that ensure LoS between the UAV and all UE. Considering the scale of the environment and the positions of obstacles, it is possible that not all UE will have LoS with the UAV, even if there is overall coverage over the entire venue. For example, if \(UE_i\) is not in LoS with the UAV, it may still be served but at a lower data rate than required, resulting in a link that fails to meet traffic demands. In such cases, deploying multiple UAVs would be necessary to meet the traffic demands of all UE and ensure LoS for each one, a scenario that is out of the scope of this work. 

\begin{figure}
	\centering
		\includegraphics[width=\linewidth]{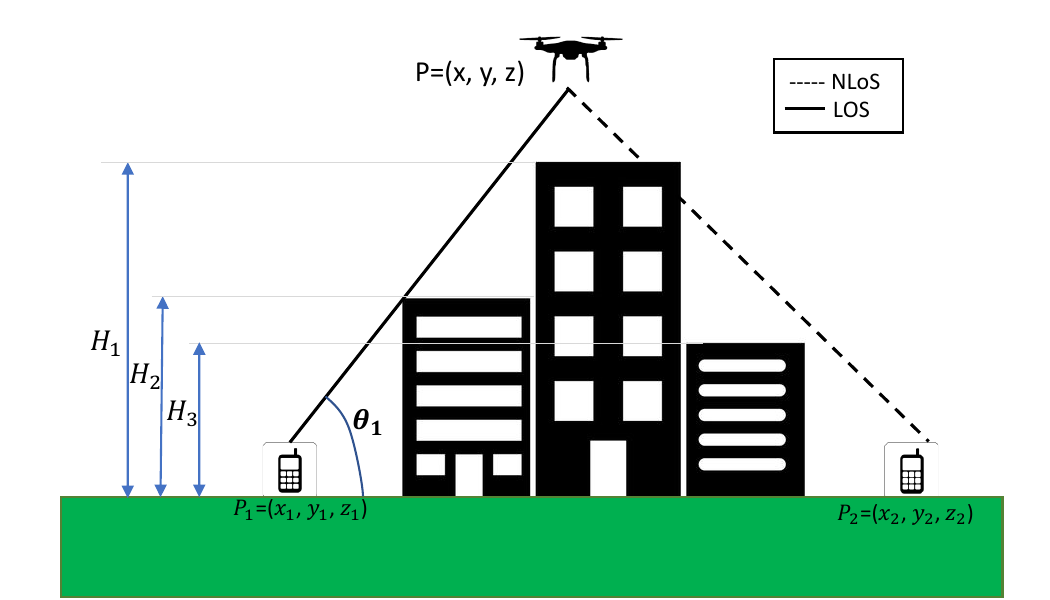}
	\caption{LoS and NLoS link representation based on ITU recommendation.}
	\label{fig.3}
\end{figure}

By implementing the ITU model and using the same mathematical formulation as our previous work (TOPA) \cite{shafafi2023joint}, we can easily determine the presence or absence of LoS between the UAV and UE. This takes into account the distribution of buildings, their coordinates, the size of the venue, and the positions of the UAV and all UE, as illustrated in Figure \ref{fig.3}. This simple example shows the LoS establishing of $UE_1$ for the angles greater than $\theta_1$.

To prioritize LoS with all UE among the possible positions inside \( S_p \), the reward is defined based on the number of LoS links ($nLoS$) between the UAV, normalized by the total number of UE. This means the reward is maximized when \( nLoS\) equals the number of UE ($N$), leading to maximum throughput. The reward function, defined in Equation \eqref{reward}, is used to train the agent to learn the optimal policy \( \pi(Act,s) \). 

\begin{equation}\label{reward}
    \centering
    \begin{aligned}    
        &Rew[t_k] =nLoS_{norm} & &\\ 
        & \text{where}\hspace{0.5cm}~nLoS_{norm} = \frac{nLoS}{N}&\\                  
    \end{aligned}
\end{equation}

\begin{algorithm}
\DontPrintSemicolon
\label{algorithm}
\KwInput{
    \textbf{Traffic demand}~ $B_i$, in $bit/s$,
    \textbf{3D Position of $UE_i$}~$P_i = (x_i, y_i, z_i)$,
    \textbf{Positions of obstacles},        
    \textbf{Noise floor power}~$P_N$, in $dBm$,
    \textbf{Movement step size}~ $gridSize$, 
    \textbf{ Feasible zone $Z_p$ for positioning the UAV},
    \textbf{Frequency}~$f$, 
    \textbf{Transmission power}~$P_{T,i}$, in $dBm$         
} 
\textbf{Calculate a sphere for every $UE_i$}\newline
\For {$(i=1; i \leq N; i \leftarrow i+1)$}{
    \textbf{Compute the minimum $SNR_i$}\;
    \textbf{Compute $d_{max,i}$} considering Equation \eqref{eq:placement-equations-system}\;    
    
} 
\textbf{Compute $S_p$}\newline
\tcc{Intersection between the spheres around each UE as the target area to position UAV}
\textbf{Initialize the 3D position of the UAV in the center of $Z_p$}\;
$t_k = 0$\;
\While{$P \in Z_p$ \text{and} \text{not} Game over}{
    \If{$P \in S_p$} {
        $nLoS = 0$\;
        \textbf{Apply ITU-R1411 propagation loss model}\newline
        \For {$(i=1; i \leq N; i \leftarrow i+1)$}{
            
            \If {$P$ is in LoS with $UE_i$}{
                $LoS_i$ = 1\;
                $nLoS \leftarrow nLoS + 1$\;
            }
            \Else{
               $LoS_i$ = 0\;
            }
        }
       $Rew[t_k] \leftarrow \frac{nLoS}{N}$\newline
       
       $Obs[t_k] \leftarrow (P_i, P, nLoS)$~\newline
    } 
    $t_k \leftarrow t_k + 1$\;    
    $P \leftarrow$ \textbf{Apply a new action}\newline \tcc{A new action defined as a new UAV position that the agent selects from the action space based on the observations and the received reward in previous $t_k$ to move forward maximum $Rew[t_k]$}\;
}
\textbf{The agent learns to find the best strategy and trajectory to find maximum accumulated reward among the action space}\newline
\tcc{The optimal position $P$ is extracted from the corresponding observations with maximum reward}
\KwOutput{\textbf{Optimal position},~$P$}
\caption{RLTOPA}
\end{algorithm}

\subsection{RLTOPA BLOCK DIAGRAM}
\label{sec:flowchart}

\begin{figure}
	\centering
		\includegraphics[width=\linewidth]{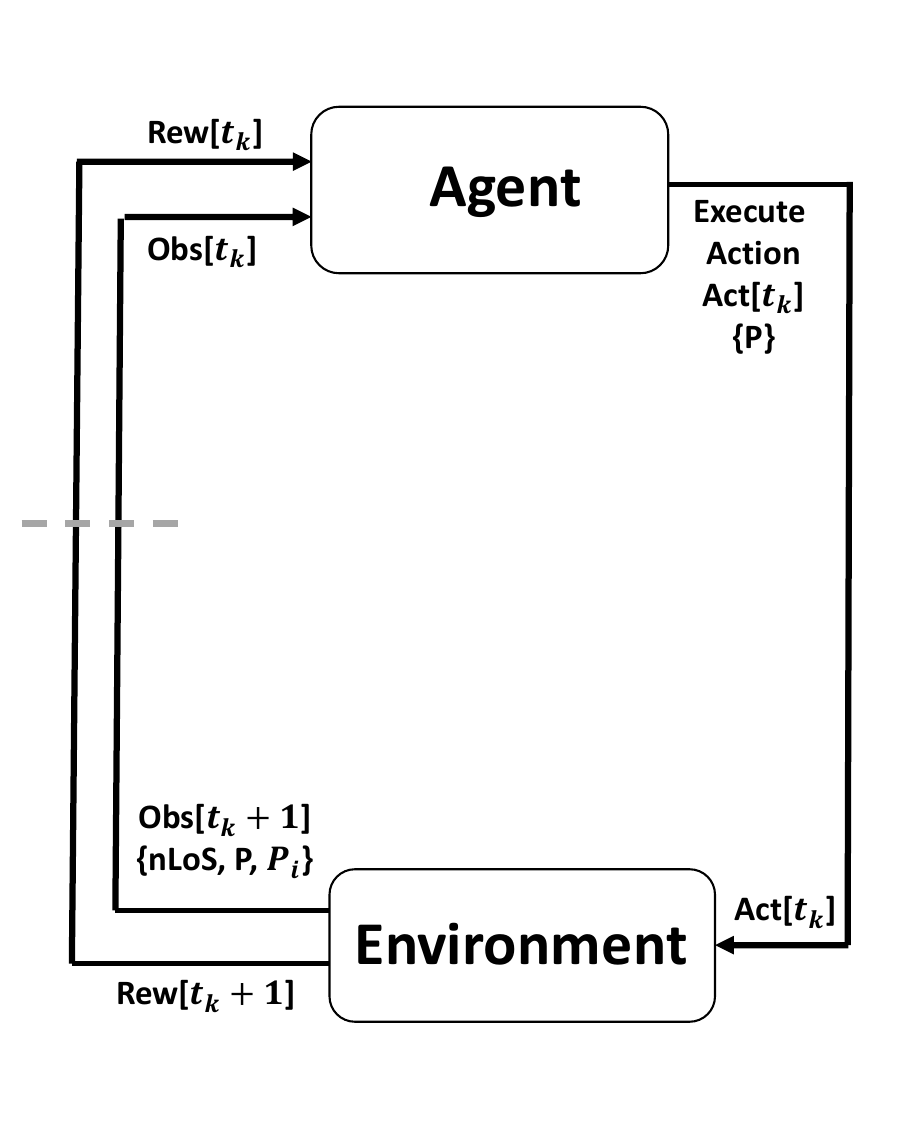}
	\caption{RLTOPA block diagram.}
	\label{fig.4}
\end{figure}

RLTOPA works using a set of input parameters, which include the coordinates of all UE, positions of obstacles, users' traffic demands, the transmission power of all UE, noise floor power, \(gridSize\), \(Z_p\), and the frequency \emph{f}. As depicted in Figure \ref{fig.4}, the environment is first established and configured based on these inputs, and the initial position of the UAV is set. Simultaneously, the sphere around each UE is calculated based on its traffic demand, and \(S_p\) is determined. 

In the first timeslot \(t_k\), the agent receives the initial observation of the current state to identify the environment's characteristics. An action involves selecting a new position for the UAV from the action space described in Section \ref{sec:action}. In each decision interval \(t_k\), the agent gathers observations from the current state, including the position of the UAV $P$, the coordinates of all UE $P_i$, \(nLoS\), and the value of the reward. Based on these observations and the reward, the agent executes an action and waits for \(t_k\) seconds. This process is repeated during each timeslot throughout the current training episode. In the subsequent episodes, based on the experience received in the previous episode, the agent plans a strategy and trajectory to move toward the maximum reward. With the increasing number of episodes, the agent learns to move toward the maximum reward faster and faster. After the agent has been trained and learned the optimal policy, RLTOPA extracts the observations associated with the maximum reward which involves the position of the UAV $P$. Algorithm \ref{algorithm} presents RLTOPA including the training process.

\section{PERFORMANCE EVALUATION}
\label{sec:PERFORMANCE EVALUATION}

\subsection{simulation setup}
\label{sec:simulation setup}

Without loss of generality, we examine an urban area with dimensions of $100m \times 100m$ under The ITU-R1411 model. The scenario includes nine buildings of varying heights by $ns3::BuildingsModule$, with a maximum building height of $20m$, and incorporates streets and alleys of different widths. RLTOPA has been implemented in ns-3 \cite{ns3adisc41:online}, a relevant tool for simulating real-world behavior in various scenarios. The environment was constructed in ns-3 to provide the foundational framework for the scenario, supplying the agent with initial and essential specifications. Additionally, RLTOPA was trained and evaluated using ns3-gym \cite{inproceedings}, which serves as an interface between the environment and the agent. By employing the ns-3 network simulation as a basis, ns3-gym facilitates the creation of an $OpenAI Gym RL$ environment and oversees the dynamic wireless networking environment employed for agent training and evaluation.

The simulation is based on Wi-Fi Access Point (AP)-Station (STA) mode. The UAV acts as the AP and the UE acts as STA. Table \ref{tab1} summarizes the parameters of ns-3.38 based on the system model presented in Section \ref{sec:SYSTEMMODEL}. As discussed in Section \ref{sec:RLTOPA}, a DQN algorithm is implemented for training the agent. Table \ref{tab2} shows the main parameters of the DQN. During the training phase, the agent begins with an epsilon-greedy factor of one, which decreases exponentially throughout the episodes. To prevent exploration during evaluation, the trained policy is loaded and deployed with a fixed epsilon-greedy factor of zero.

\subsection{simulation scenarios}
\label{sec: simulation scenarios}

RLTOPA was evaluated across various scenarios with different traffic demands and number of UE to assess its performance. We conducted two main scenarios. In Scenario A, four UE are randomly deployed. The scenario is configured according to Table \ref{tab1} to evaluate two cases. In the first case, a homogeneous setup is established where all four users have the same traffic demands. We assume that the demanded capacity for all UE is 58.5 Mbit/s, associated with the IEEE 802.11ac MCS index 0. In the second case, heterogeneous settings are considered. The traffic demands of UE are different in this case. Two settings are considered: 
i) \(B_0 = B_1 = 2 \times B_2 = 2 \times B_3\), which assumes traffic demands of 117 Mbit/s associated with the IEEE 802.11ac MCS index 1 for \(B_0\) and \(B_1\), and 58.5 Mbit/s associated with MCS index 0 for \(B_2\) and \(B_3\);
ii) \(B_0 = 0.75 \times B_1 = 2 \times B_2 = 4 \times B_3\). Traffic demands of 234, 175.5, 117, and 58.5 Mbit/s associated with IEEE 802.11ac MCS index 3, 2, 1, and 0 are considered for \(B_0\), \(B_1\), \(B_2\), and \(B_3\) respectively.

Scenario B is introduced to assess the performance of RLTOPA in more demanding and congested environments. In Scenario B, twelve UE are deployed, and multiple settings similar to Scenario A are explored. A traffic demand of 58.5 Mbit/s is applied for a homogeneous setting, while in a heterogeneous setup, configurations from Scenario A are applied to groups of UE. Specifically, the initial case encompasses two distinct traffic demands for every six UE, whereas the other case introduces four distinct traffic demands, each associated with three users.

During the training phase, RLTOPA is executed for 10 episodes in each scenario with the same state and settings outlined in Table \ref{tab2}. The agent is trained to maximize the reward by utilizing the information received from the state. The agent aims to position itself at 3D coordinates with $nLoS$ equal to the number of UE. The performance of RLTOPA is evaluated in all scenarios, using two key metrics: i) aggregate throughput received by the UAV; and ii) delay, which is the average time, including queuing, transmission, and propagation delays, taken for the packets to arrive at the UAV from the instant they are generated by the source application of each UE.

 \begin{table}[ht]
    \caption{\textbf{ NS-3 environment configuration parameters for RLTOPA.}}
    \label{tab1}
    \setlength{\tabcolsep}{3pt}
    \begin{tabular}{p{120pt} p{100pt}}
        \hline
        \multicolumn{2}{c}{ns-3.38 simulation parameters} \\  
        \hline
        $S_{venue}$ & 100 m  \\        
        $N$ & variable with the scenario\\
        $f$ & 5250 MHz\\
        $M$& 9 $^{\mathrm{a}}$ \\
        Guard Interval $(GI)$ & 800 $ns$\\
        Wi-Fi channel & 50 \\
        Wi-Fi Standard & IEEE 802.11ac  \\
        Channel Bandwidth & 20 MHz  \\
        Antenna Gain & 0 dBi  \\
        Tx power& 20 $dBm$\\
        Noise floor power & -85 $dBm$\\
        LoS Propagation Loss Model & ItuR1411LosPropagationLossModel   \\
        NLoS Propagation Loss Model & ItuR1411NlosOverRooftopP ropagationLossModel  \\
        Remote Station Manager mechanism & IdealWifiManager  \\
        Application Traffic & UDP constant bitrate \\
        UDP Data Rate $(B_i)$ & variable based on $MCS_i$ \\
        Packet Size & 1400 bytes  \\
        \hline
        \multicolumn{2}{c}{Positioning zone ($Z_p$) Parameters} \\  
        \hline
        $(X_{min}, Y_{min}, Z_{min})$ & (-50.0, -50.0, 25)\\
        $(X_{max}, Y_{max}, Z_{max})$ & (50.0, 50.0, 100)\\
        girdSize& 1 m \\
        \hline
        \multicolumn{2}{c}{Buildings Coordinates} \\
        \multicolumn{2}{c}{$(x_{min}, x_{max}, y_{min}, y_{max}, z_{min}, z_{max}, floors, x_{rooms}, y_{rooms})$}\\
        \hline
        \multicolumn{2}{c}{$(-5.0, 5.0, -5.0, 5.0, 0.0, 20.0, 5, 3, 2)$}\\
        \multicolumn{2}{c}{$(-5.0, 5.0, 20.0, 30.0, 0.0, 15.0, 4, 3, 2)$}\\
        \multicolumn{2}{c}{$(-5.0, 5.0, -30.0, -20.0, 0.0, 15.0, 4, 3, 2)$}\\
        \multicolumn{2}{c}{$(-35.0, -25.0, -5.0, 5.0, 0.0, 20.0, 5, 3, 2)$}\\
        \multicolumn{2}{c}{$(-35.0, -25.0, 20.0, 30.0, 0.0, 20.0, 5, 3, 2)$}\\
        \multicolumn{2}{c}{$(-35.0, -25.0, -30.0, -20.0, 0.0, 15.0, 4, 3, 2)$}\\
        \multicolumn{2}{c}{$(25.0, 35.0, -5.0, 5.0, 0.0, 20.0, 5, 3, 2)$}\\
        \multicolumn{2}{c}{$(25.0, 35.0, 20.0, 30.0, 0.0, 15.0, 4, 3, 2)$}\\
        \multicolumn{2}{c}{$(25.0, 35.0, -30.0, -20.0, 0.0, 15.0, 4, 3, 2)$}\\            
        \hline           
        \multicolumn{2}{p{220pt}}{$^{\mathrm{a}}$ The number of buildings can be adjusted based on the venue volume; however, for simplicity and without loss of generality, we consider nine buildings.}
    \end{tabular}
\end{table}

\begin{table}[ht]
    \caption{\textbf{Summary of DQN learning algorithm parameters.}}
    \label{tab2}
    \setlength{\tabcolsep}{3pt}
    \begin{tabular}{p{100pt} p{120pt}}
        \hline
        \multicolumn{2}{c}{DQN Learning Algorithm Parameters} \\  
        \hline    
        Number of episodes & 10\\
        Duration of the episodes& 300 s\\
        Decision interval, $t_k$ & 100 ms\\
        Start time of training & 2.1 s \\
        Observation Space & five-dimensional scaled float \\
        Action Space & one-dimensional discrete scaled integer \\
        ML library & TensorFlow\\
        Optimizer & Adam (learning rate of $10^{-2}$) \\
        Epsilon Greedy & 1 (random decision) $^{\mathrm{a}}$ \\ 
        Quadratic Loss & Mean Square Error (MSE) \\
        Q-function& Two completely linked layers, each with 32 units\\
        Memory Replay & buffer size is $10^6$ with a batch of 64\\
        \hline
        \multicolumn{2}{p{220pt}}{$^{\mathrm{a}}$ The rate falls through polynomial decay towards 0.1 (10\%) random decisions.}
    \end{tabular}
\end{table}
\begin{figure*}[!t]
    \centering    
    \subfloat[Reward for episodes 1,5, and 10\label{fig:5.1}]{
        \includegraphics[scale=0.28]{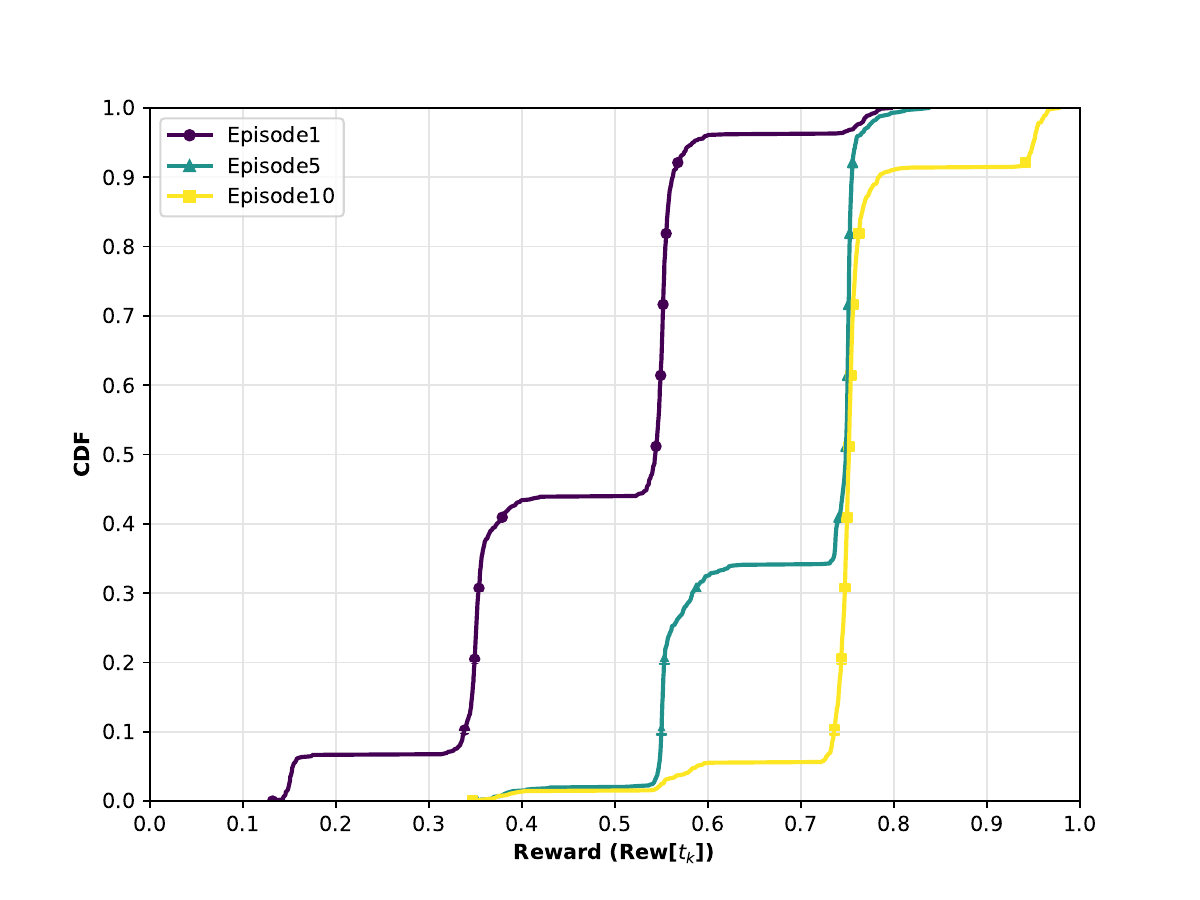}        
    } 
    \hfill
    \subfloat[Aggregate throughput (Mbit/s)\label{fig:5.2}]{       
        \includegraphics[scale=0.28]{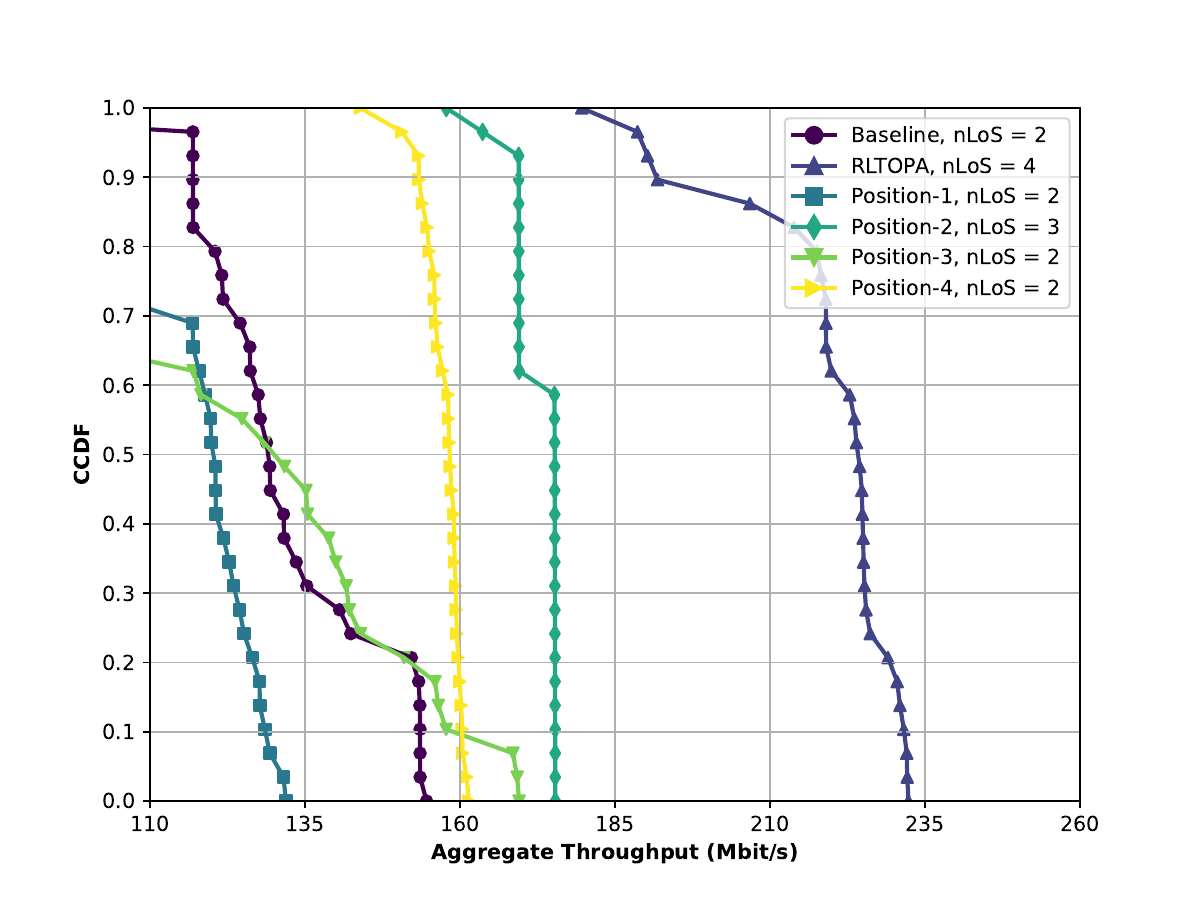}        
    }   
    \hfill
    \subfloat[Delay (s)\label{fig:5.3}]{       
        \includegraphics[scale=0.28]{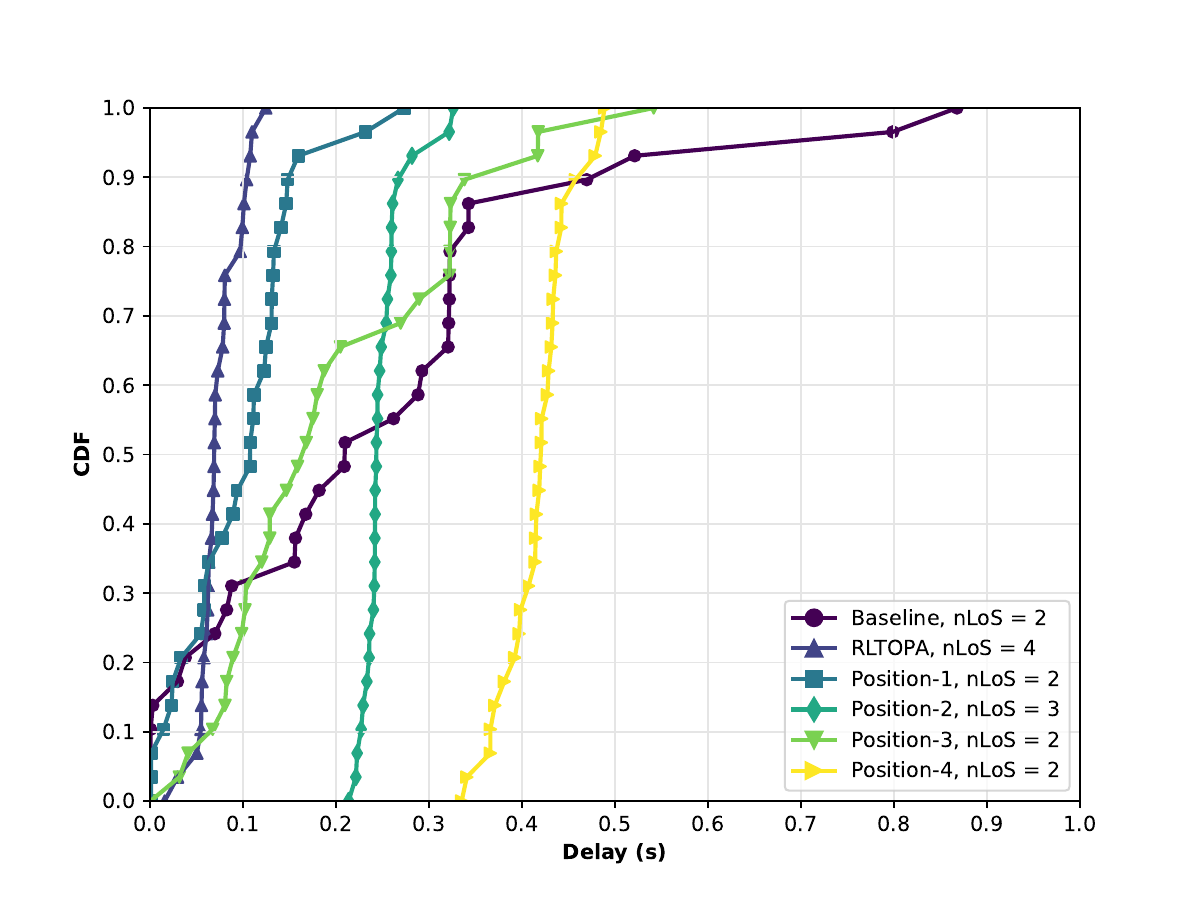}        
    }
    \caption{Scenario A -- four UE with homogeneous traffic demands: $B_0 = B_1 = B_2 = B_3$.}
    \label{fig5}
\end{figure*}
\begin{figure*}[!t]
    \centering    
    \subfloat[Reward for episodes 1,5, and 10\label{fig:6.1}]{
        \includegraphics[scale=0.28]{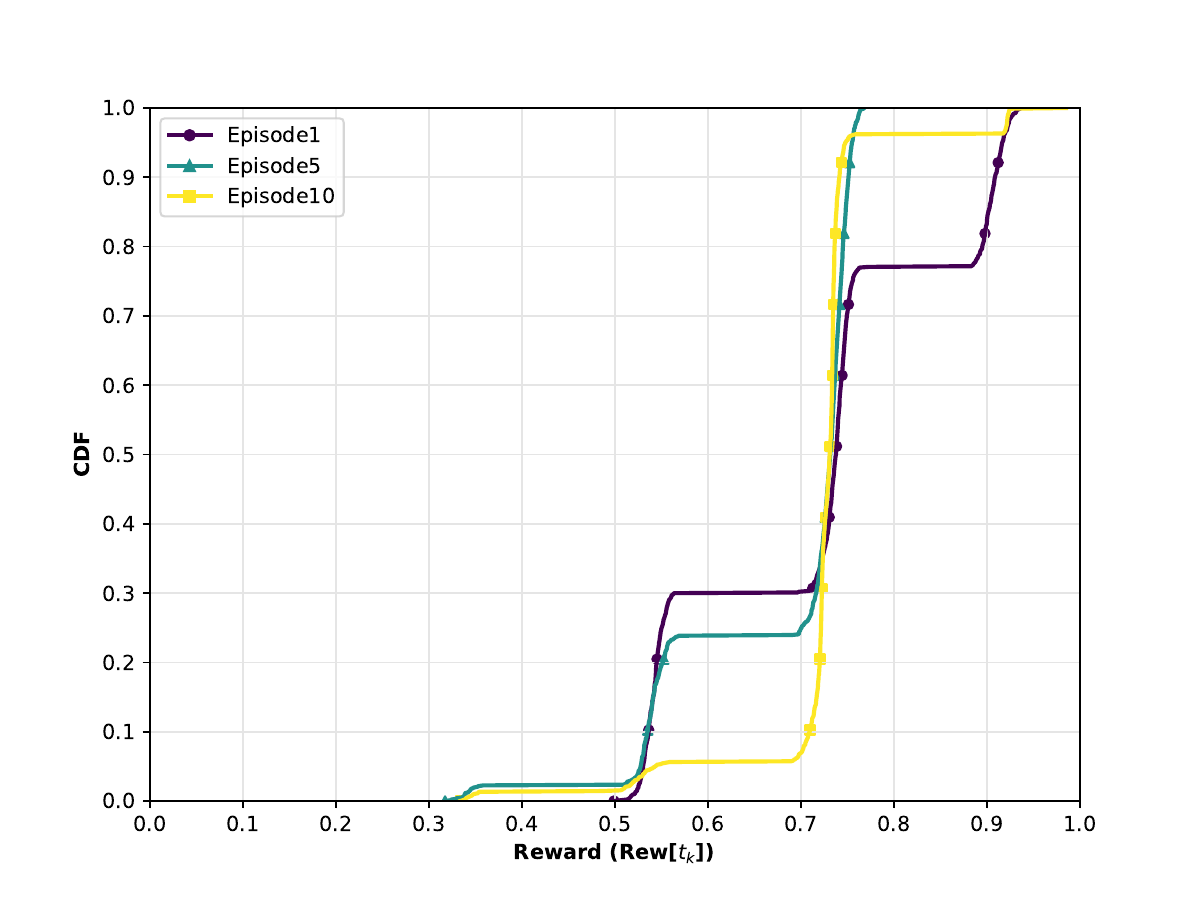}        
    } 
    \hfill
    \subfloat[Aggregate throughput (Mbit/s)\label{fig:6.2}]{       
        \includegraphics[scale=0.28]{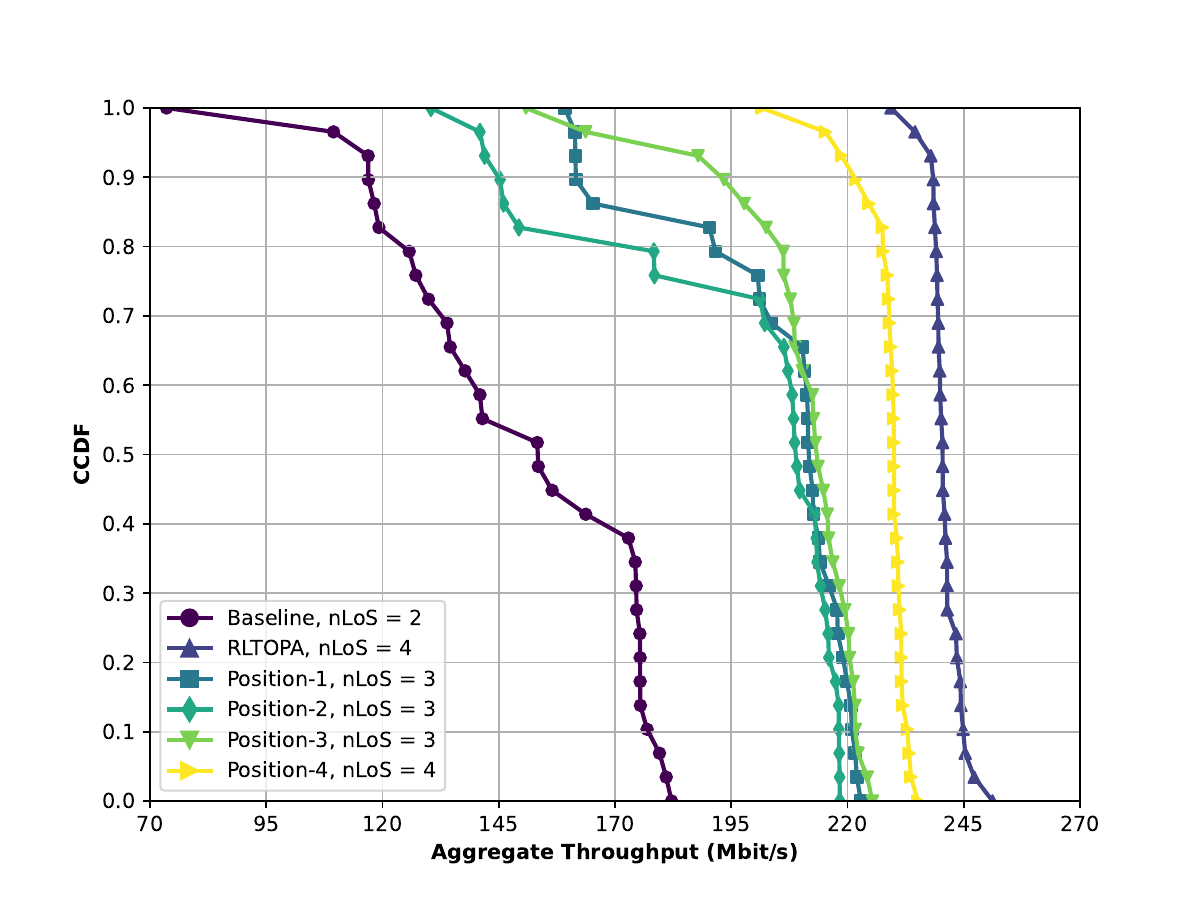}        
    }   
    \hfill
    \subfloat[Delay (s)\label{fig:6.3}]{       
        \includegraphics[scale=0.28]{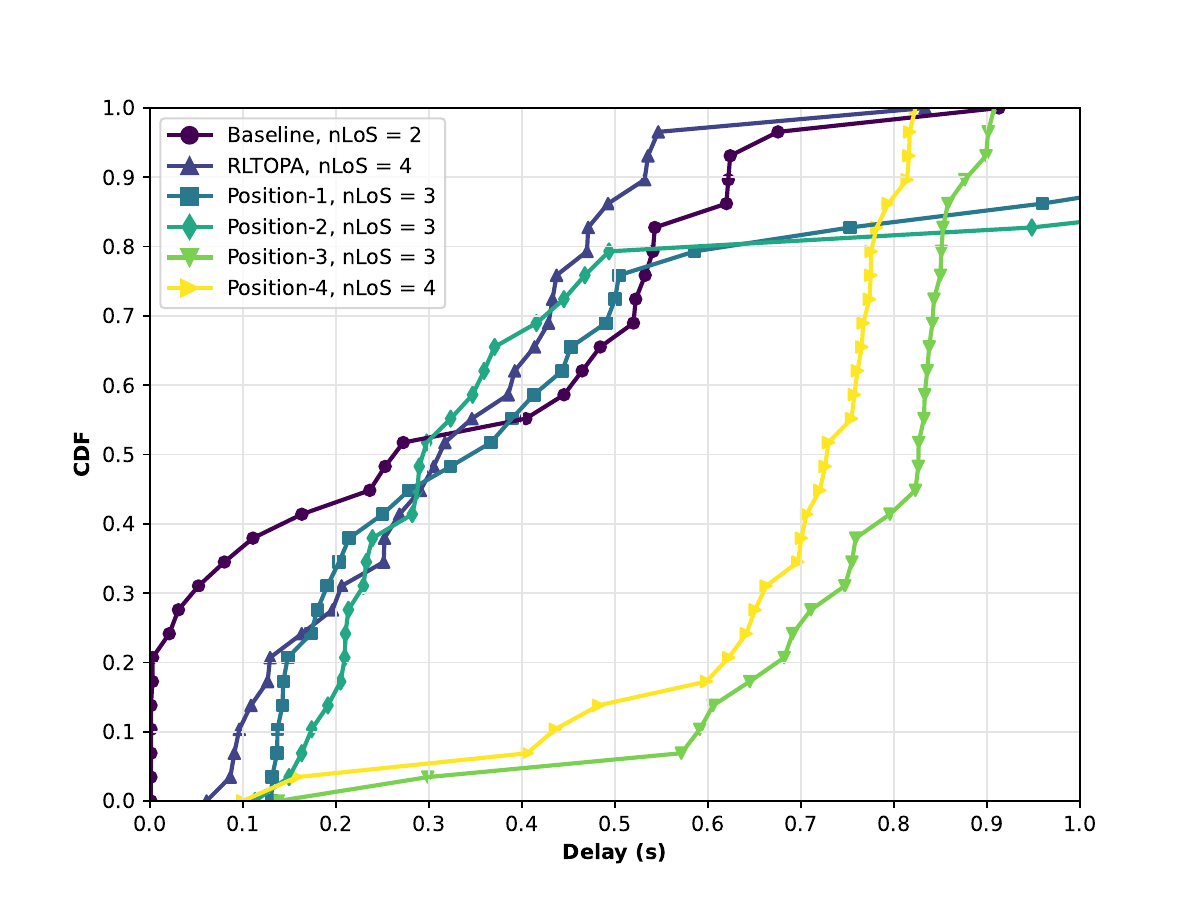}        
    }
    \caption{Scenario A -- four UE with heterogeneous traffic demands: $B_0 = B_1 = 2 \times B_2 = 2 \times B_3$.}
    \label{fig:6}
\end{figure*}
\begin{figure*}[!t]
    \centering    
    \subfloat[Reward for episodes 1,5, and 10\label{fig:7.1}]{
        \includegraphics[scale=0.28]{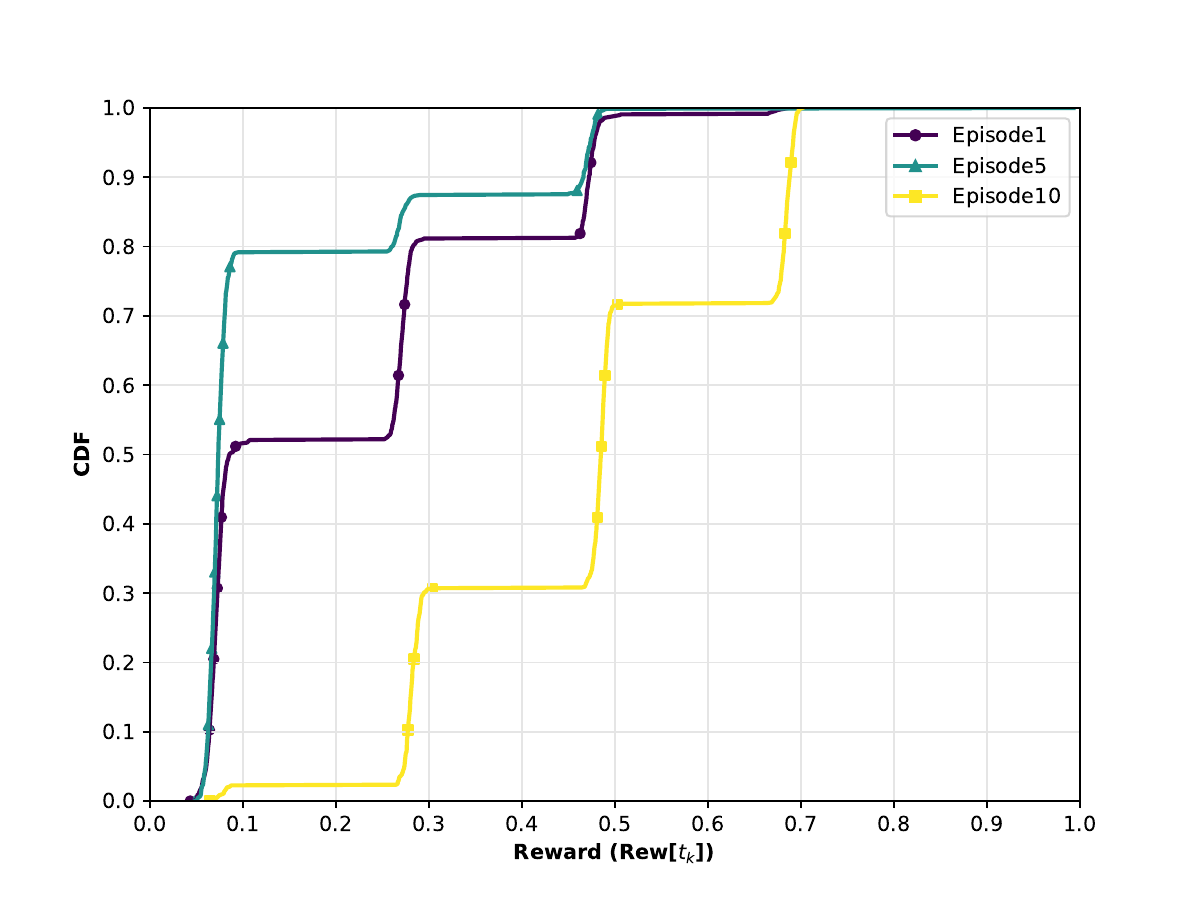}        
    } 
    \hfill
    \subfloat[Aggregate throughput (Mbit/s)\label{fig:7.2}]{       
        \includegraphics[scale=0.28]{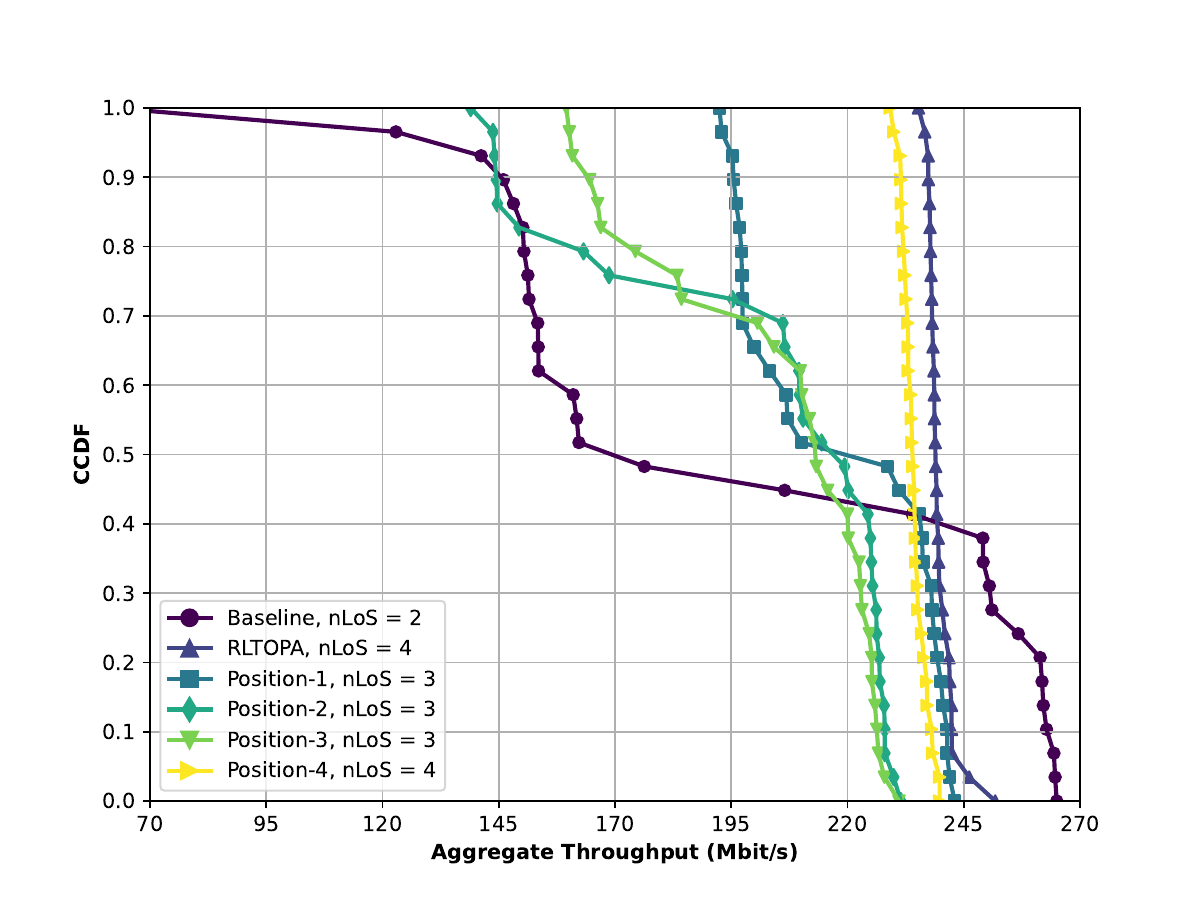}        
    }   
    \hfill
    \subfloat[Delay (s)\label{fig:7.3}]{       
        \includegraphics[scale=0.28]{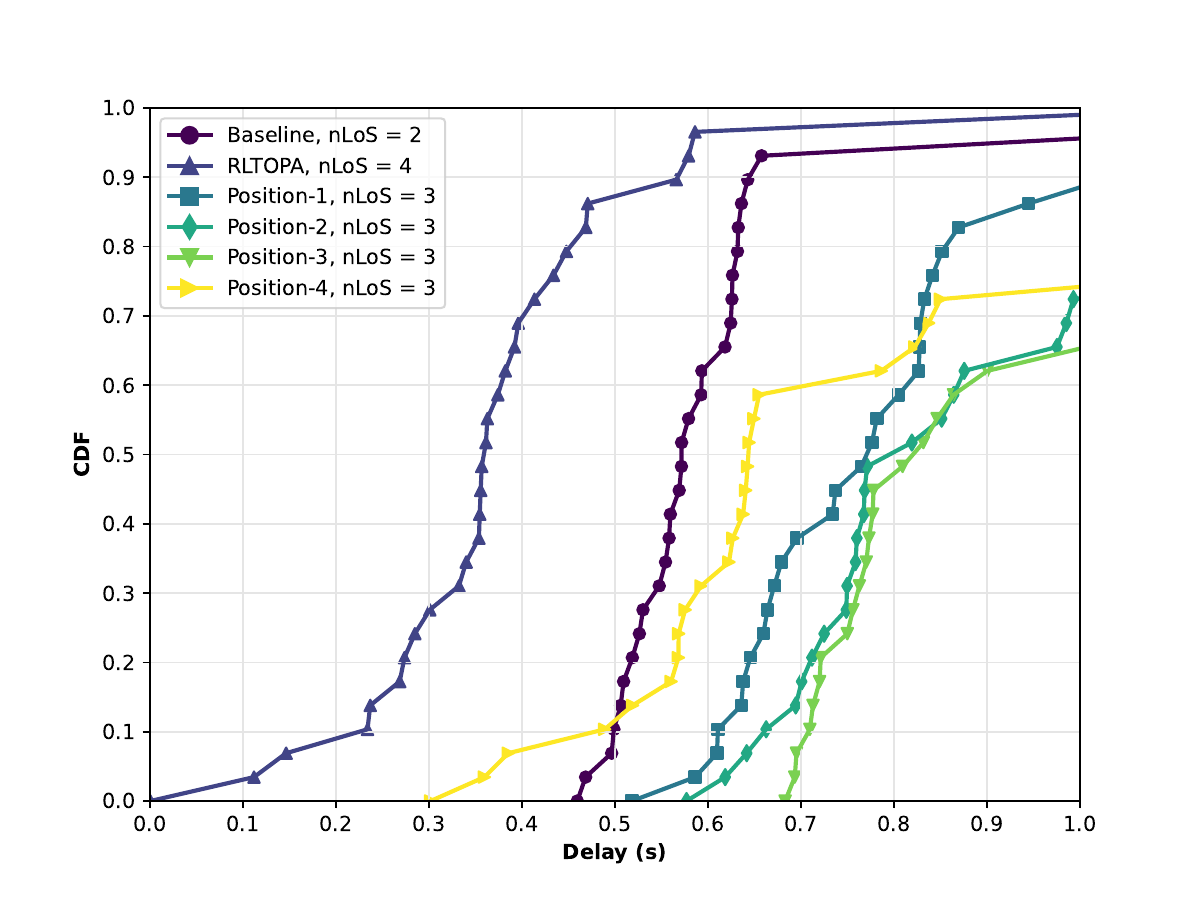}        
    }
    \caption{Scenario A -- four UE with heterogeneous traffic demands: \(B_0 = 0.75 \times B_1 = 2 \times B_2 = 4 \times B_3\).}
    \label{fig:7}
\end{figure*}
\subsection{Simulation Results}
\label{sec: Simulation Results}  
This section presents the simulation results. The training results are derived from 10 episodes, each lasting 300s, while evaluation results are obtained from 30 simulation runs for each scenario outlined in Section \ref{sec: simulation scenarios}. All simulations were conducted under consistent networking conditions, utilizing \(SetRandomSeed()\) and \(RngRun = \{1, 2, ..., 30\}\). The results are visualized using the complementary cumulative distribution function (CCDF) to depict the distribution of aggregate throughput received by the UAV. The cumulative distribution function (CDF) is used to present the reward and the mean delay. The results highlight the optimal position achieved by RLTOPA and compare it with five alternative positions in each scenario. The baseline position is a stationary AP positioned at the center of the venue, mounted on the rooftop of the central building. The remaining positions, denoted as \(Position-i\), where \(i \in \{1, ..., 4\}\), are situated 10m away from the optimal position, respectively, in the left, right, front, and behind directions.

\begin{figure*}[!t]
    \centering    
    \subfloat[Reward for episodes 1,5, and 10\label{fig:8.1}]{
        \includegraphics[scale=0.28]{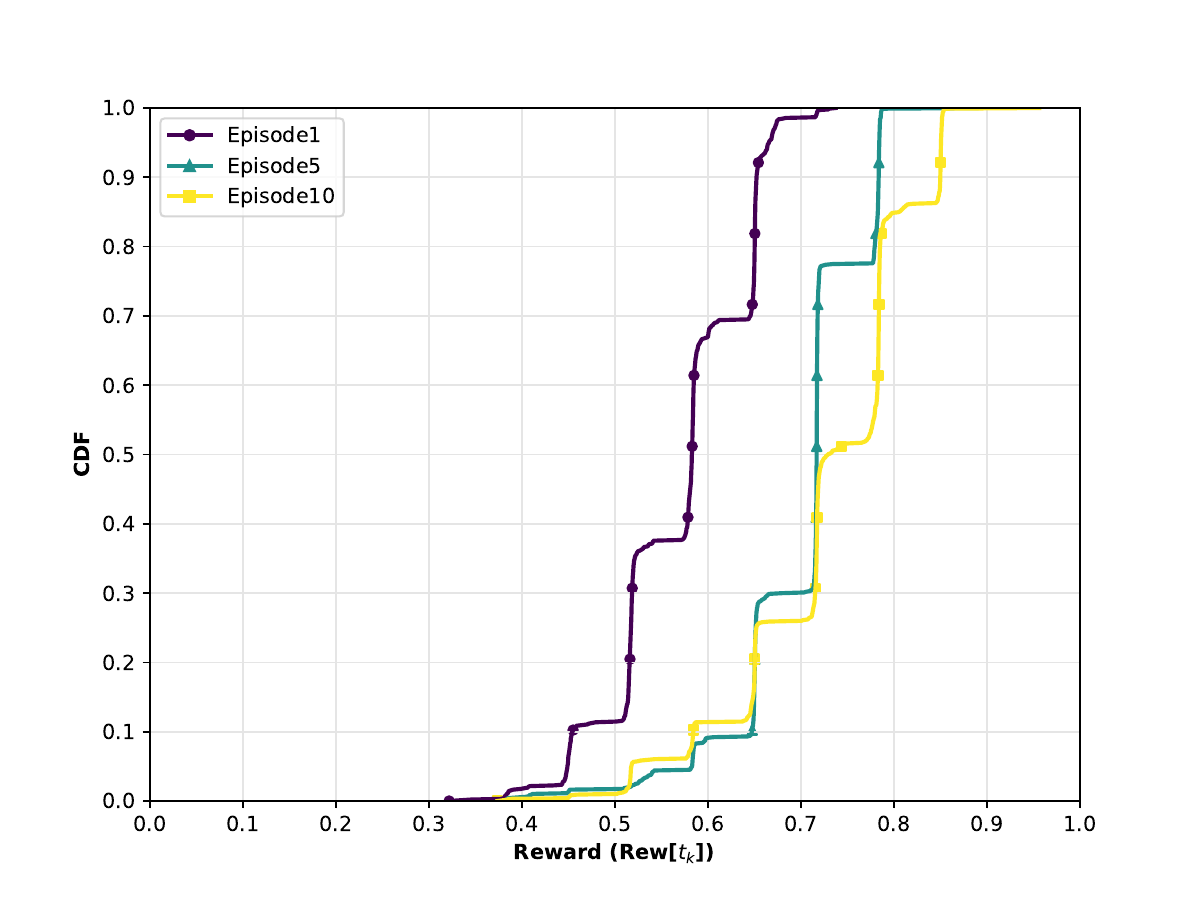}        
    } 
    \hfill
    \subfloat[Aggregate throughput (Mbit/s)\label{fig:8.2}]{       
        \includegraphics[scale=0.28]{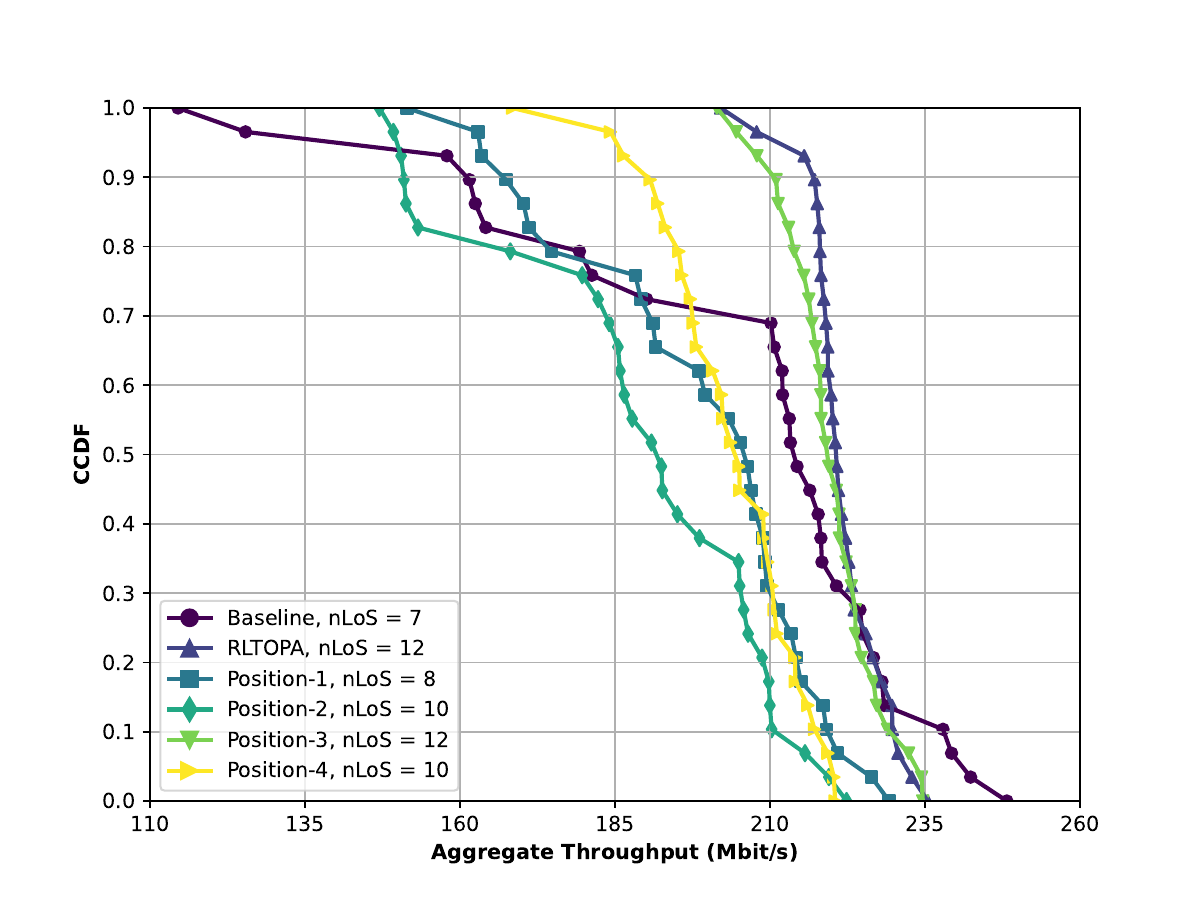}        
    }   
    \hfill
    \subfloat[Delay (s)\label{fig:8.3}]{       
        \includegraphics[scale=0.28]{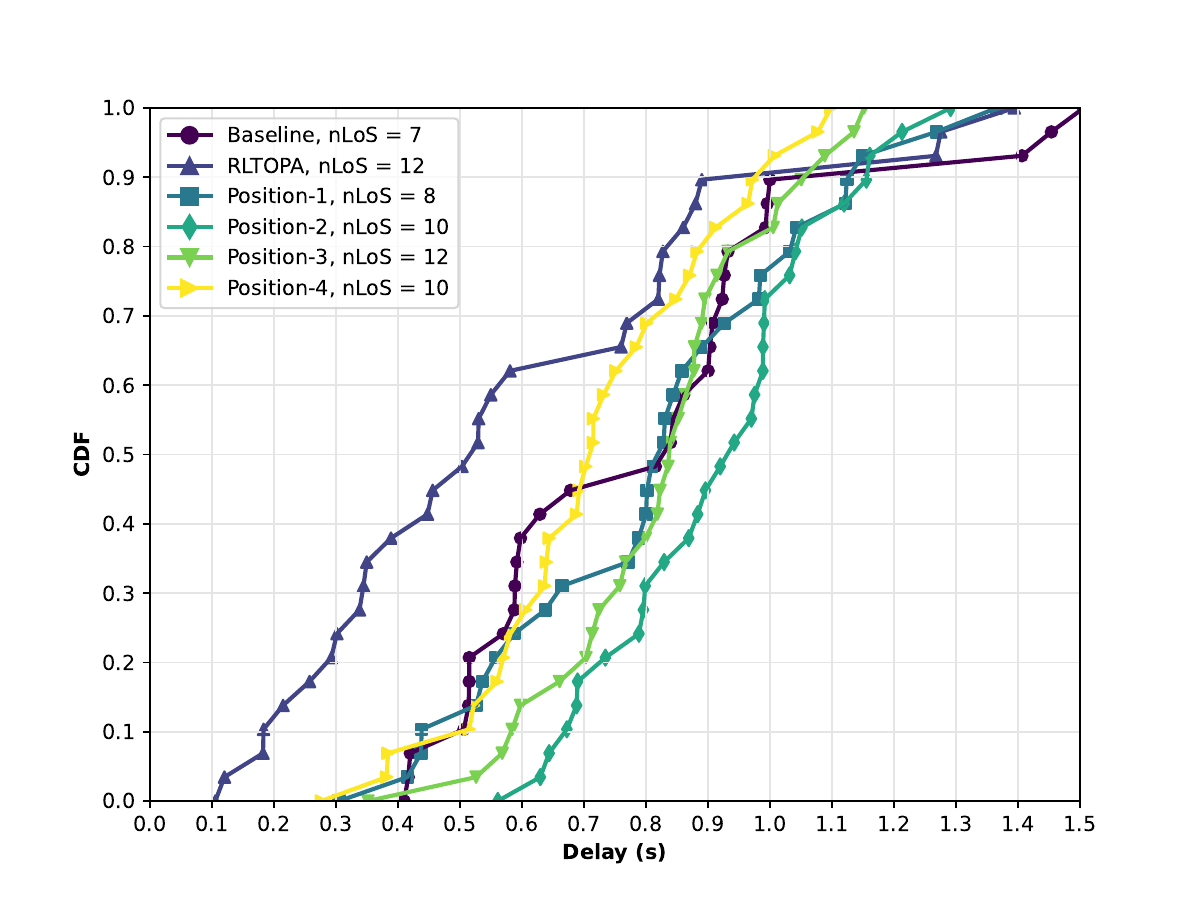}        
    }
    \caption{Scenario B -- twelve UE with homogeneous traffic demands: $B_0 = B_1 = B_2 = B_3$. Each traffic demand is associated with three UE.}
    \label{fig:8}
\end{figure*}
\begin{figure*}[!t]
    \centering    
    \subfloat[Reward for episodes 1,5, and 10\label{fig:9.1}]{
        \includegraphics[scale=0.28]{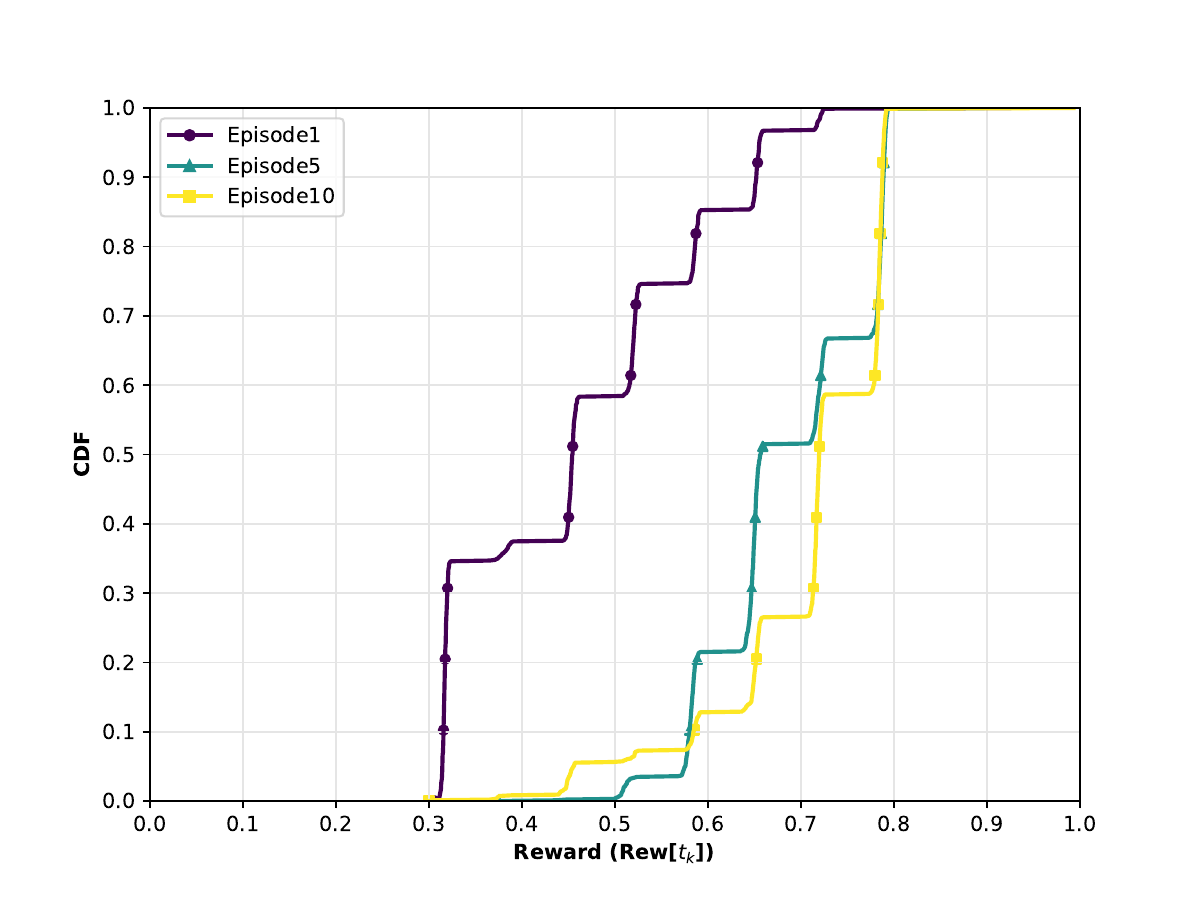}        
    } 
    \hfill
    \subfloat[Aggregate throughput (Mbit/s)\label{fig:9.2}]{       
        \includegraphics[scale=0.28]{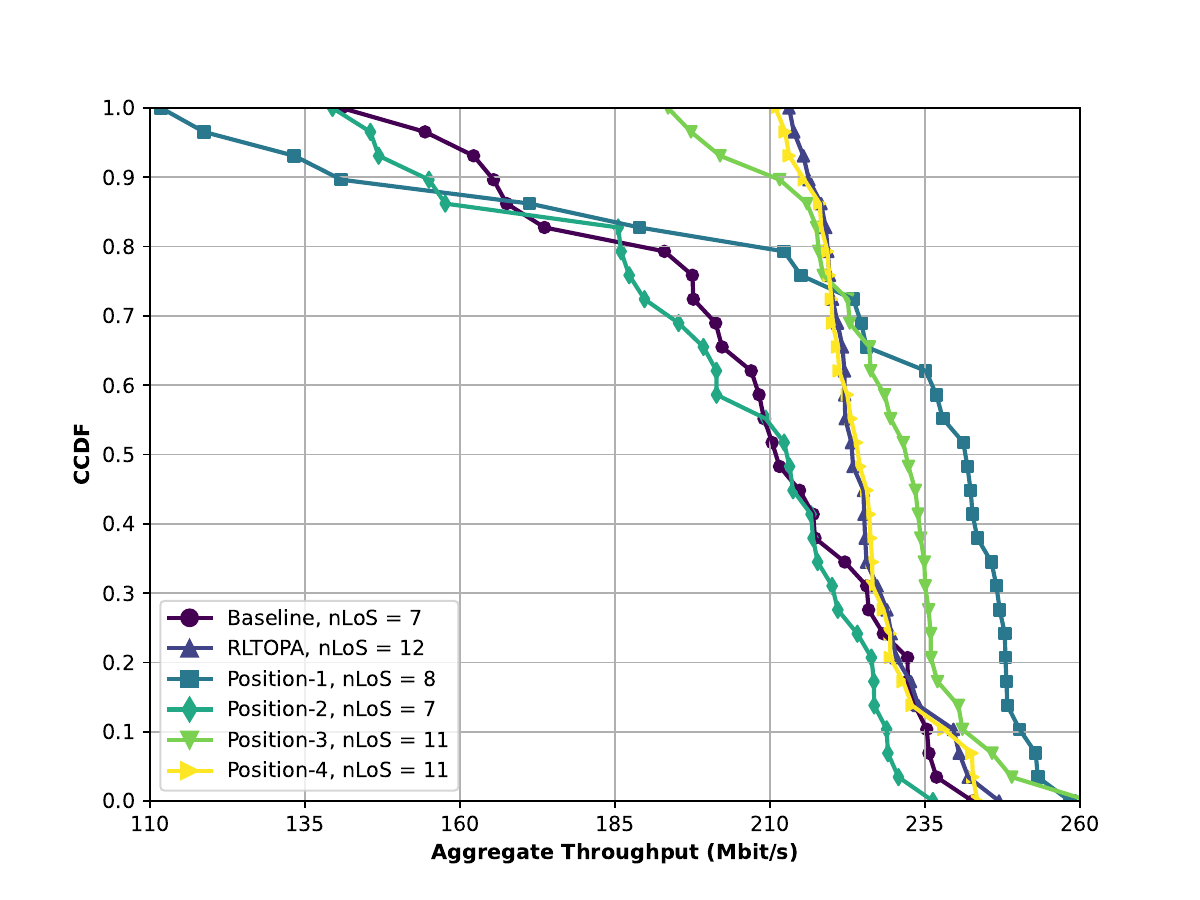}        
    }   
    \hfill
    \subfloat[Delay (s)\label{fig:9.3}]{    
        \includegraphics[scale=0.28]{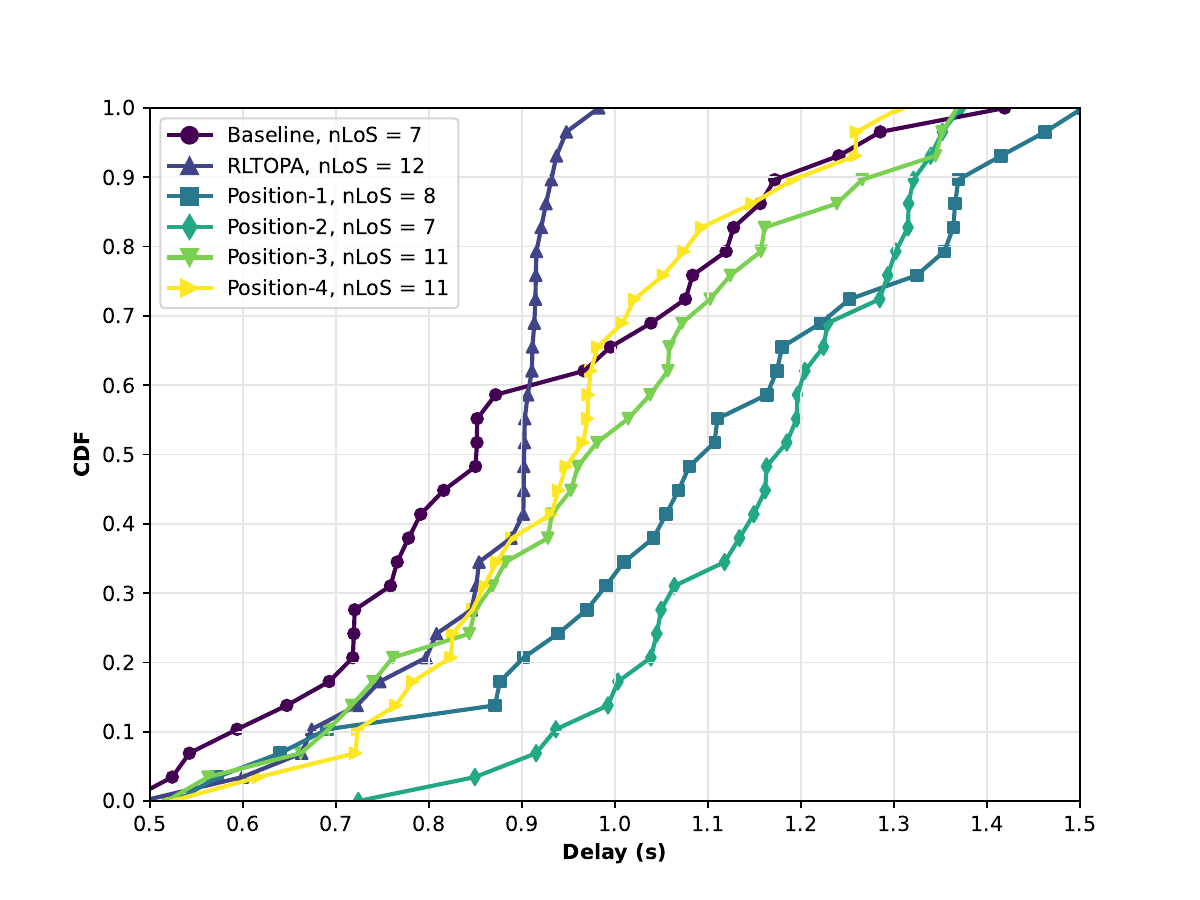}        
    }
    \caption{Scenario B -- twelve UE with heterogeneous traffic demands: $B_0 = B_1 = 2 \times B_2 = 2 \times B_3$. Each traffic demand is associated with three UE.}
    \label{fig:9}
\end{figure*}
\begin{figure*}[!t]
    \centering    
    \subfloat[Reward for episodes 1,5, and 10\label{fig:10.1}]{
        \includegraphics[scale=0.28]{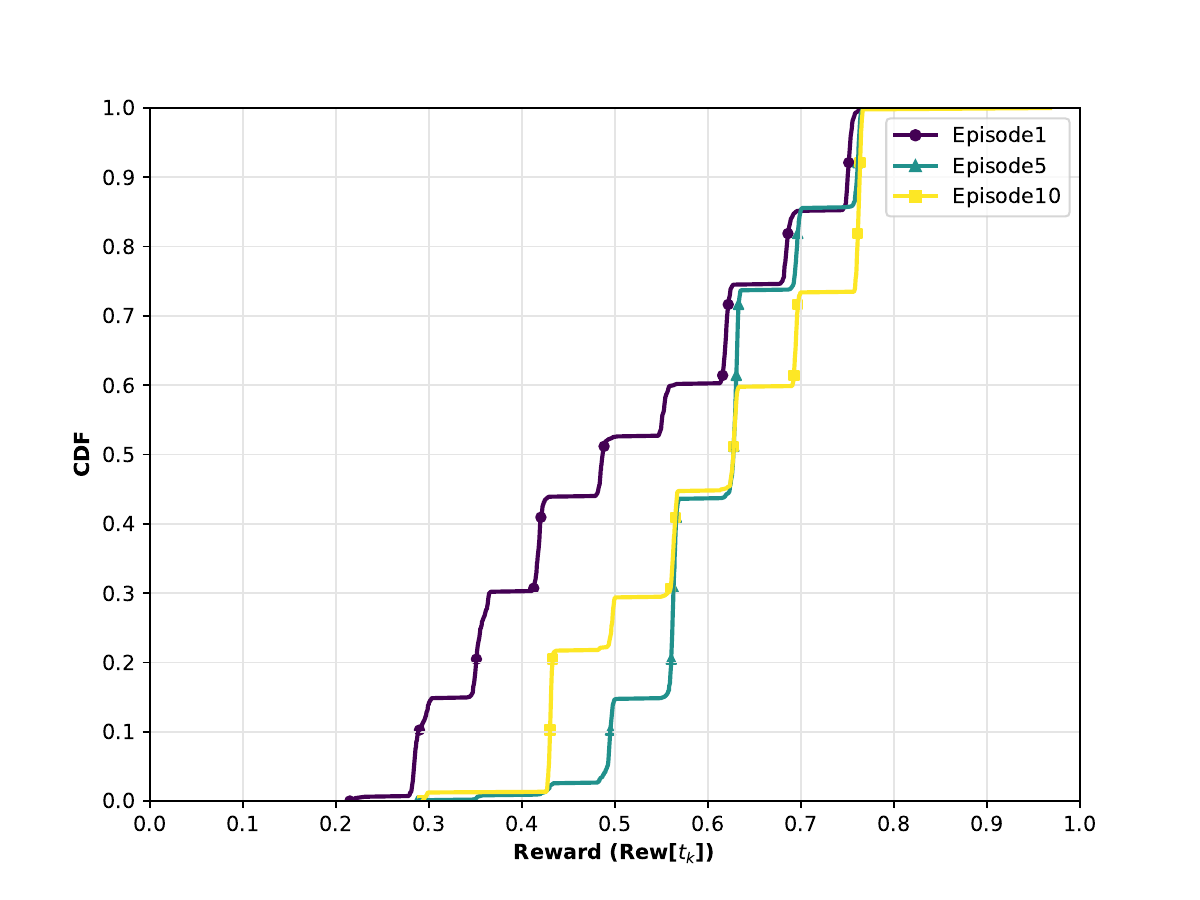}        
    } 
    \hfill
    \subfloat[Aggregate throughput (Mbit/s)\label{fig:10.2}]{       
        \includegraphics[scale=0.28]{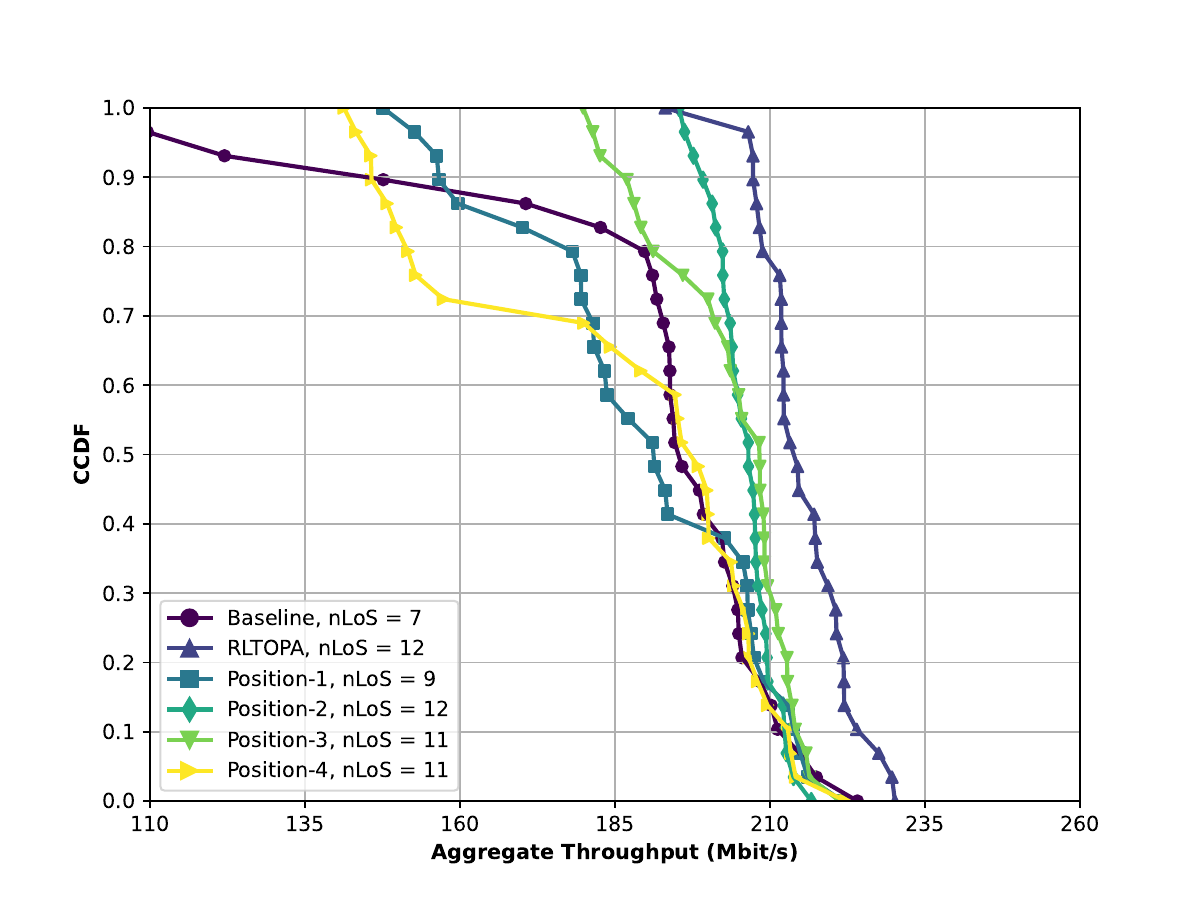}        
    }   
    \hfill
    \subfloat[Delay (s)\label{fig:10.3}]{       
        \includegraphics[scale=0.28]{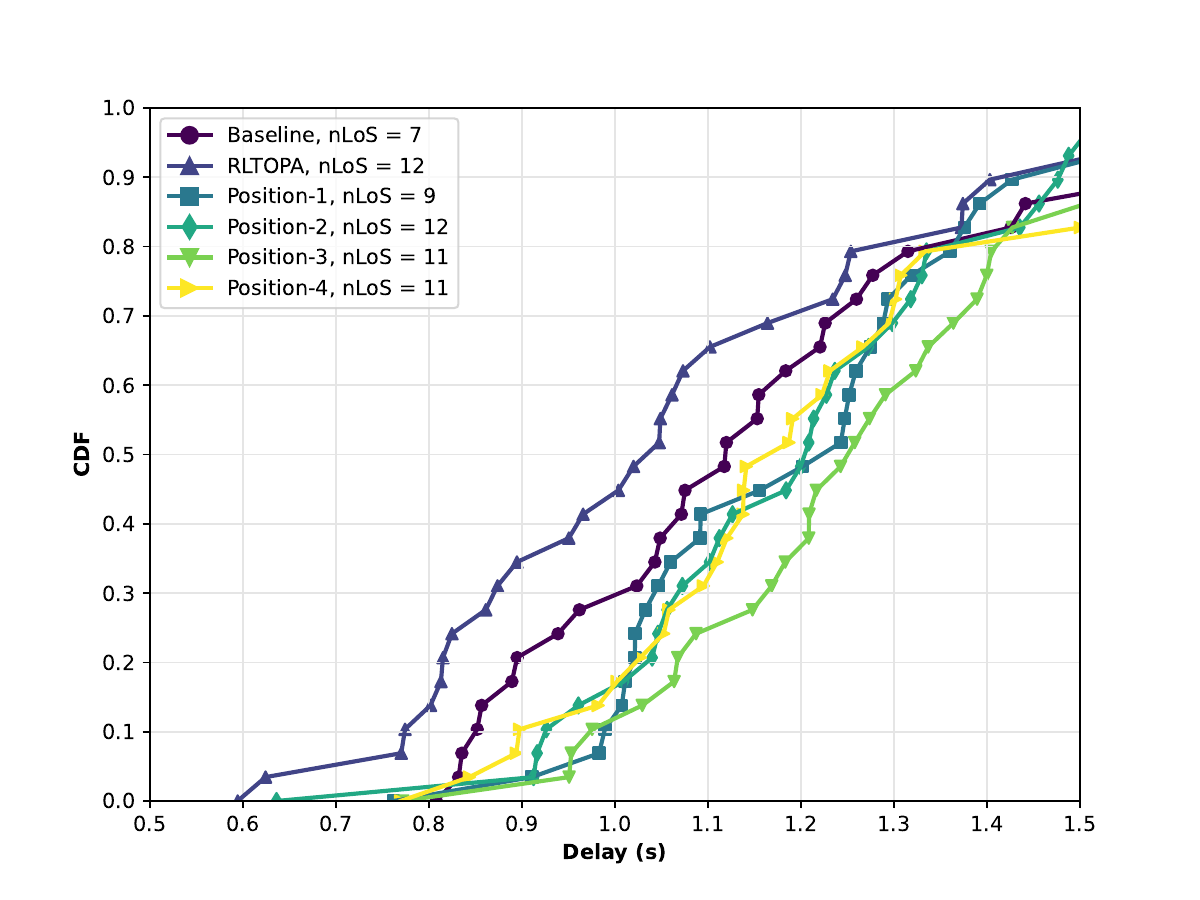}        
    }
    \caption{Scenario B -- twelve UE with the heterogeneous traffic demands: \(B_0 = 0.75 \times B_1 = 2 \times B_2 = 4 \times B_3\). Each traffic demand is associated with three UE.}
    \label{fig:10}
\end{figure*}
Figure \ref{fig5} represents Scenario A, deploying four UE with the same traffic demands as stated in Section \ref{sec: simulation scenarios}. Figure \ref{fig:5.1} shows the CDF presentation of the reward function for episodes 1, 5, and 10. From this figure, we can observe that the agent successfully moved toward the maximum reward in all episodes, achieving this more quickly in the later episodes, and the adopted training process enabled the last episode to have a better median (0.75) than the first episode (0.55). The 3D optimal position of the UAV achieved by RLTOPA with maximum nLoS is (-28, 18, 33) in this scenario. Figures \ref{fig:5.2} and \ref{fig:5.3} show the evaluation of the RLTOPA in terms of aggregate throughput and mean delay. RLTOPA leads to significant improvements in aggregate throughput and mean delay. Specifically, compared to the baseline position, the aggregate throughput is enhanced up to 58\% for the 90th percentile and up to 69\% for the 50th percentile. The mean delay decreased by more than 71\% in the 50th percentile compared to the baseline. Position-5 has $nLoS = 4$.

With the same configuration and the same number of UE in identical positions, but with varying traffic demands, the optimal position identified by RLTOPA is (-15, -26, 30). Notably, even when the UE's positions remain unchanged, the optimal position of the UAV differs from the previous scenario. This variation is due to the need to maximize both nLoS and throughput, considering the arrangement of user traffic demands. This highlights the performance and efficiency of RLTOPA. As depicted in Figure \ref{fig:6.1}, RLTOPA achieves both the maximum reward and maximum nLoS. Figure \ref{fig:6.2} shows a 95\% improvement in throughput compared to the baseline for the 90th percentile and a 74\% improvement for the 50th percentile. Additionally, the delay was reduced by 21\% for the 50th percentile, as depicted in Figure \ref{fig:6.3}.

In the final case of Scenario A, RLTOPA identifies the position (-21, 22, 30) to provide LoS for all users while maximizing throughput. The plot of Figure \ref{fig:7.1} shows the agent strives to maximize the reward to determine the optimal UAV position. In this heterogeneous scenario, RLTOPA boosts aggregate throughput by 50\% for the 50th percentile and 65\% for the 90th percentile. The delay decreases by 49\% for the 90th percentile, as illustrated in Figures \ref{fig:7.2} and \ref{fig:7.3}. 

To evaluate RLTOPA under more challenging and crowded conditions and assess its performance with varying numbers of UE, Scenario B was designed. The agent trains and explores for optimal rewards, as depicted in Figure \ref{fig:8.1}. Twelve users are distributed in the venue with the same configuration as in Scenario A, with all UE having identical traffic demands, as stated in Section \ref{sec: simulation scenarios}. After executing RLTOPA, the position (-41, -2, 37) was identified as optimal to serve all users with LoS. The CCDF of the aggregate throughput is shown in Figure \ref{fig:8.2}. In this scenario, RLTOPA achieves a 4\% improvement in throughput for the 50th percentile and a 38\% improvement for the 90th percentile. Figure \ref{fig:8.3} shows the delay is reduced by 37\% for the 50th percentile.

Similar to Scenario A, we considered two distinct traffic demands to assess the compatibility of RLTOPA with heterogeneous scenarios. Figures \ref{fig:9} and \ref{fig:10} display the results for these cases. Figure \ref{fig:8.1} illustrates the CDF of rewards for the first case, with traffic demands as described in Section \ref{sec: simulation scenarios}. The optimal position identified is (-33, 0, 30), ensuring maximum LoS and throughput. Figure \ref{fig:9.2} indicates a 17\% improvement for the 90th percentile, a 5\% improvement for the 50th percentile, and a 19\% reduction in delay for the 90th percentile (Figure \ref{fig:9.3}). In the last case, RLTOPA identified the position (-27, -12, 51) as optimal for the UAV, ensuring LoS for all UE. A 30\% improvement in throughput for the 90th percentile and a 13\% improvement for the 50th percentile are depicted in Figure \ref{fig:10.2}. The delay has been reduced by 9\%, as shown in Figure \ref{fig:10.3}. 

\begin{figure}
	\centering
		\includegraphics[width=\linewidth]{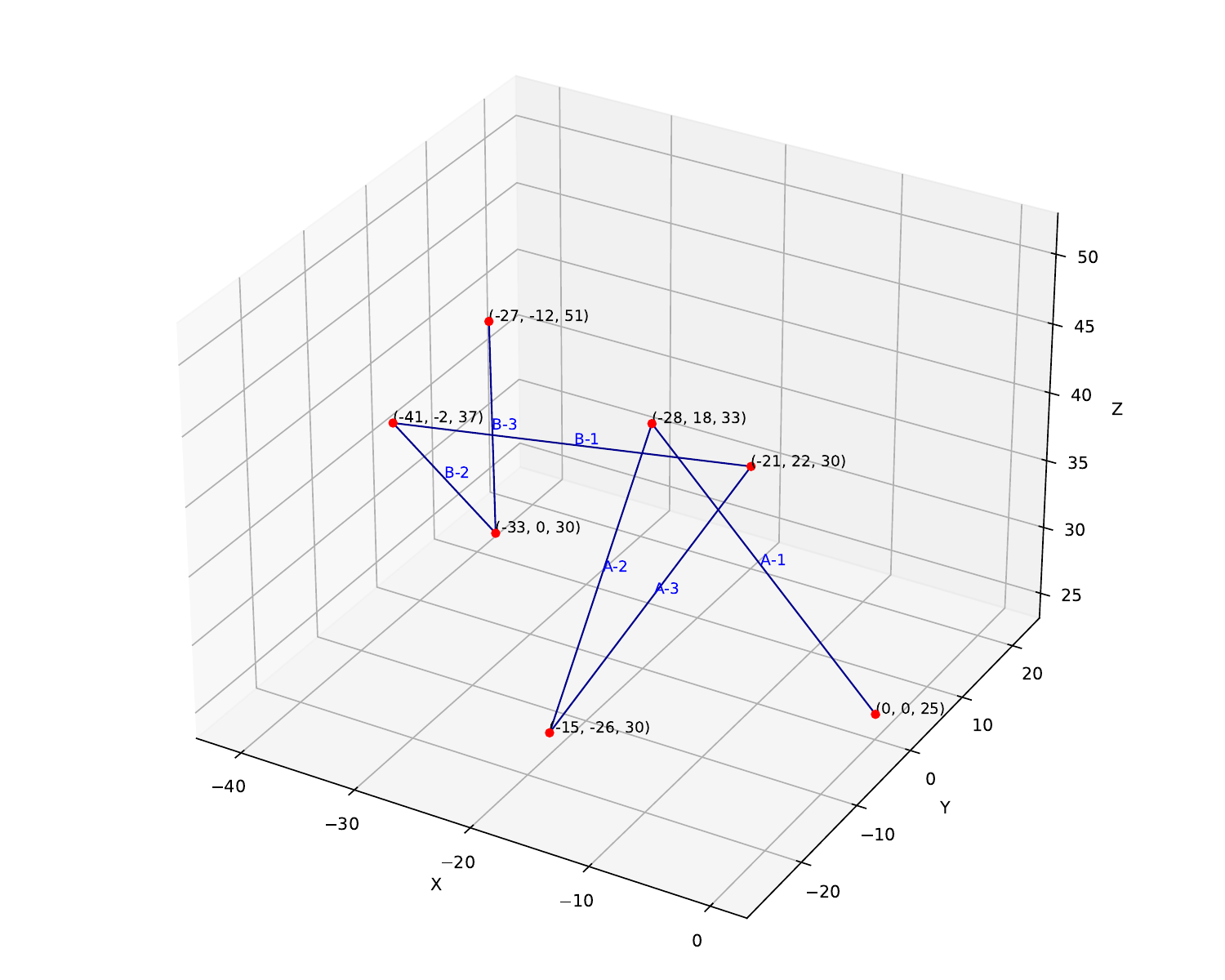}
	\caption{Optimal positions achieved by RLTOPA in each scenario}
	\label{fig.12}
\end{figure}

The results illustrate the performance of RLTOPA across scenarios characterized by different numbers of UE, traffic demands, and UE distribution. In all simulated scenarios, RLTOPA was able to determine the optimal position for the UAV. Figure \ref{fig.12} illustrates traffic-awareness of RLTOPA. While the positions of the UE remain constant between cases in each scenario, the UAV's optimal position varies due to changes in traffic demands.


\section{conclusions}
\label{sec: conclusions}

In this paper, we introduced RLTOPA, a novel algorithm designed for real-world scenarios that leverages RL. RLTOPA considers the positions of all UE, obstacles within the venue, and users' traffic demands to determine the optimal UAV position. The aim is to accommodate traffic demands by providing LoS between the UAV and all UEs. We conducted various experiments and scenarios to evaluate RLTOPA's performance under different configurations and combinations of user traffic demands. Additionally, we demonstrated RLTOPA's compatibility with varying numbers of UE and both homogeneous and heterogeneous traffic demands. 

The results show up to a 95\% improvement in aggregate throughput and a 71\% improvement in delay without compromising fairness. For future work, enhancing RLTOPA to address dynamic scenarios could be beneficial. Additionally, deploying multiple UAVs in larger-scale scenarios could offer a promising solution for extended coverage. Also, leveraging signal processing \cite{5381162} and computer vision to detect the position of the UE and obstacles can be part of future work.

\section*{Acknowledgment}
This work is financed by National Funds through the Portuguese funding agency, FCT – Fundação para a Ciência e a Tecnologia, under the PhD grant 2023.00384.BD.

\bibliographystyle{IEEEtran}
\bibliography{ref}

\begin{IEEEbiography}[{\includegraphics[width=1in,height=1.25in,clip,keepaspectratio]{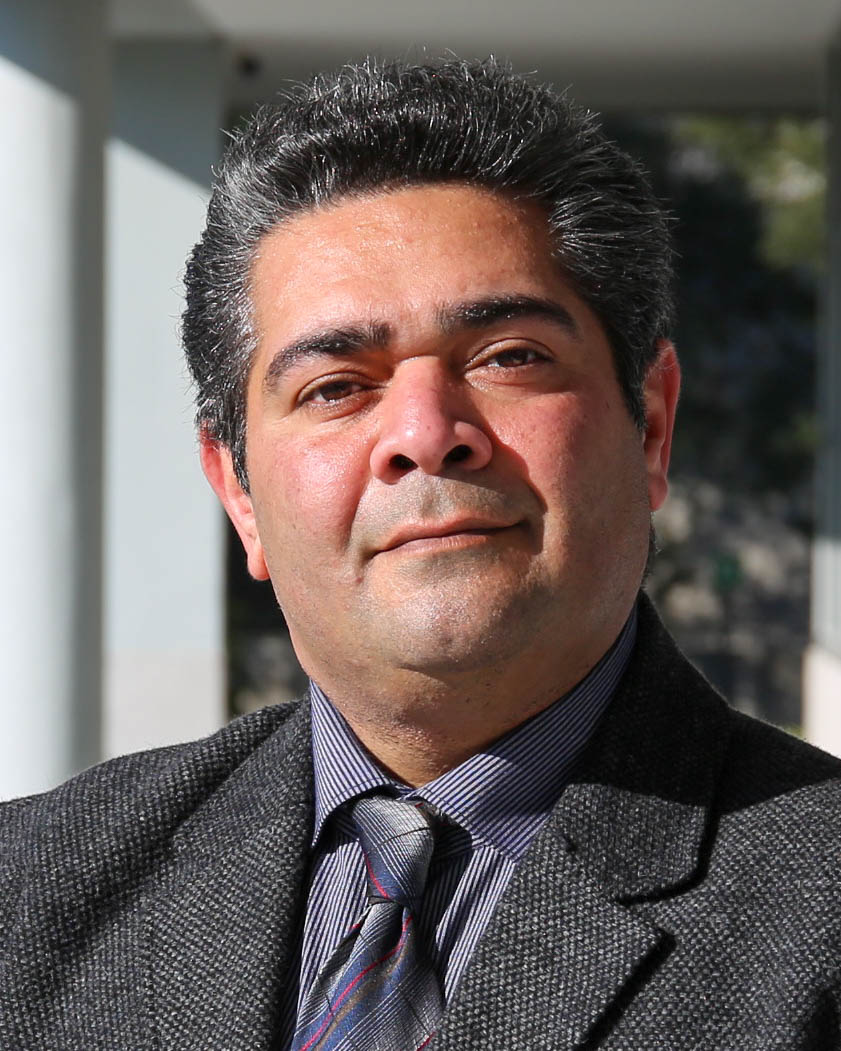}}]{Kamran Shafafi} received the M.Sc. degree in telecommunication engineering in 2008. Currently, he is pursuing a Ph.D. degree in telecommunications at the University of Porto. He is a researcher at INESC TEC and a trusted reviewer for several journals and conferences such as IEEE ACCESS and Elsevier, with more than 80 paper reviews. Since 2008, he has been an invited professor at several universities. He is the author of nine books and more than 20 national and international articles. His research interests include aerial networks and vision-based positioning for flying networks. He has participated in more than 15 national and international research and development projects, such as INEFECT, ResponDrone, and CONVERGE.

\end{IEEEbiography}

\begin{IEEEbiography}[{\includegraphics[width=1in,height=1.25in,clip,keepaspectratio]{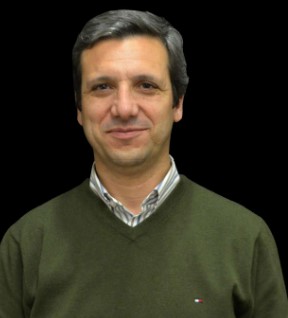}}]{Manuel Ricardo}  (Member, IEEE) received a Ph.D. degree in Electrical
and Computer Engineering (EEC) from the Faculty of Engineering of the University of Porto (FEUP) in 2000. He is
a Full Professor with FEUP, where he
teaches courses on mobile communications and
computer networks.  He is the Director of the Department of Electrical and Computer Engineering of FEUP, and Associate Director of the INESC TEC research institute. His research interests include
mobile communications networks, quality of service, radio resource management, and performance
assessment.
\end{IEEEbiography}

\begin{IEEEbiography}[{\includegraphics[width=1in,height=1.25in,clip,keepaspectratio]{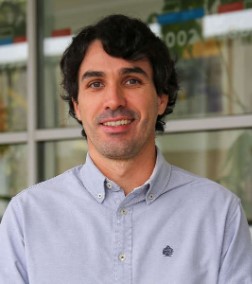}}]{Rui Campos} (Senior Member, IEEE) received
the Ph.D. degree in electrical and computers
engineering from the University of Porto, in 2011.
He is currently a Senior Researcher and the
Coordinator of the Centre for Telecommunications
and Multimedia with INESC TEC and also an
Assistant Professor with the University of Porto,
where he teaches courses on telecommunications.
He has coordinated several research projects
including the WiFIX project in CONFINE Open
Call 1, SIMBED (Fed4FIRE+ OC3), and SIMBED+ (Fed4FIRE+ OC5).
He has also participated in several EU research projects including H2020
ResponDrone, H2020 RAWFIE, FP7 SUNNY, and FP6 Ambient Networks.
He is the author of more than 85 scientific publications in international
conferences and journals with peer review. His research interests include
airborne, maritime, and green networks, with a special focus on medium
access control, radio resource management, and network autoconfiguration.
He has been a TPC Member of several international conferences, including
IEEE INFOCOM, IEEE ISCC, and IFIP/IEEE Wireless Days. He was the
General Chair of Wireless Days 2017 and the TPC Vice-Chair of 2021 Joint
EuCNC \& 6G Summit.

\end{IEEEbiography}

\

\EOD

\end{document}